\documentclass[floats,aps,epsf,amsfonts]{revtex4}
\usepackage[all]{xy}
\usepackage{amsmath}
\usepackage[dvips]{graphicx}

\begin{document}


\title{Frequency-domain calculation of the self force: \\
The high-frequency problem and its resolution}

\author{Leor Barack$^{1}$, Amos Ori$^{2}$
and Norichika Sago$^{1}$}
\affiliation
{$^1$School of Mathematics, University of Southampton, Southampton,
SO17 1BJ, United Kingdom\\
$^2$Department of Physics, Technion---Israel Institute of Technology,
Haifa, 32000, Israel}

\date{\today}

\begin{abstract}

The mode-sum method provides a practical means for calculating the self force
acting on a small particle orbiting a larger black hole. In this method,
one first computes the spherical-harmonic $l$-mode contributions $F^{\mu}_l$ of
the ``full force'' field $F^{\mu}$, evaluated at the particle's location, and then sums
over $l$ subject to a certain regularization procedure. In the frequency-domain
variant of this scheme the quantities $F^{\mu}_l$ are obtained by fully
decomposing the particle's self field into Fourier-harmonic modes $l m \omega$,
calculating the contribution of each such mode to $F^{\mu}_l$, and then summing over
$\omega$ and $m$ for given $l$. This procedure has the advantage that one only
encounters {\it ordinary} differential equations. However, for eccentric
orbits, the sum over $\omega$ is found to converge badly at the particle's
location. This problem (reminiscent of the familiar Gibbs phenomenon of Fourier
analysis) results from the discontinuity of the time-domain $F^{\mu}_l$ field at the
particle's worldline. Here we propose a simple and practical method to resolve
this problem. The method utilizes the {\it homogeneous} modes $l m \omega$ of
the self field to construct $F^{\mu}_l$ (rather than the inhomogeneous modes, as in
the standard method), which guarantees an exponentially-fast convergence to the
correct value of $F^{\mu}_l$, even at the particle's location. We illustrate the
application of the method with the example of the monopole scalar-field
perturbation from a scalar charge in an eccentric orbit around a Schwarzschild
black hole. Our method, however, should be applicable to a wider range of
problems, including the calculation of the gravitational self-force using
either Teukolsky's formalism, or a direct integration of the metric
perturbation equations.

\end{abstract}

\maketitle

\section{Introduction}

The problem of calculating the gravitational self-force
\cite{MST,QW,GW} acting on a pointlike test
particle as it moves in orbit around a black hole is attracting considerable
attention in recent years \cite{Poisson:2003nc,Lousto:2005an}.
Within this context, various authors have been studying also the analogous
problem of the scalar-field self-force \cite{Q} acting on a particle endowed
with a scalar charge, which proved to be a useful toy model.
The electromagnetic (EM) self-force, acting on an electric point charge,
was also studied by various authors
following the seminal work by DeWitt and Brehme \cite{DeWittBrehme}.
A practical algorithm, commonly used for computing the self force in all
three cases (gravitational, EM and scalar), is the {\it mode-sum method}
\cite{Barack:1999wf,Barack:2001gx,Barack:2002mh}.
This method requires as input the multipole modes of the full (retarded)
perturbation fields, along with their derivatives, evaluated at the particle's
location. These multipoles can be calculated using either frequency-domain
methods (as in, e.g., \cite{Burko:2000xx,Detweiler:2002gi,Keidl:2006}) or time-domain
numerical evolution (as in, e.g., \cite{Barack:2000zq,Barack:2002ku,Haas:2007kz,
Barack:2007tm}). In both computational approaches one sets off by writing down
the appropriate set of perturbation equations, modeling the source term
associated with the point particle (namely, the energy-momentum, the electric
four-current, or the scalar charge) as a delta-function distribution. In the
frequency-domain approach one then decomposes the inhomogeneous perturbation
equations into Fourier-harmonic modes (``$lm\omega$ modes'') and proceeds by
solving the resulting ordinary differential equations (ODEs) with suitable
boundary conditions at spatial infinity and at the event horizon. In the
alternative, time-domain approach, one refrains from decomposing the field into
frequency modes, and instead tackles the partial differential equations
for each $l,m$ directly using time evolution.

Each of the above two approaches has its strengths and weaknesses.
The time-domain approach has the advantage that one only
deals with a single field for each $l m$, whereas in the frequency-domain approach
one has to sum over the various $\omega$ modes. On the other hand, the latter
approach has the obvious advantage that one only faces ODEs.
Despite the fact that time-domain methods are winning growing popularity in
recent years, frequency-domain calculations remain an appealing option for some
range of orbital parameters \cite{Barton}.
Also, it turns out that the non-radiative multipoles of the gravitational
perturbation in Schwarzschild (i.e., the modes $l=0,1$) are difficult to analyze
in the time domain, due to instabilities \cite{Barack:2002ku,Barack:2007tm},
and one resorts to a frequency-domain calculation in this case \cite{BGSprep}.
Working in the frequency domain, however, brings about a technical
issue which, to the best of our knowledge, has not been addressed so far in
the current context.

To illustrate the problem, is it instructive to refer to the simple case
of (minimally-coupled and massless) scalar-field perturbations from a pointlike
scalar charge orbiting a Schwarzschild black hole.
In this case, the scalar field $\Phi(t,r,\theta,\varphi)$
can be decomposed in spherical harmonics $Y_{lm}(\theta,\varphi)$,
yielding the multipolar mode functions $\phi_{lm}(t,r)$.
Here $t,r,\theta,\varphi$ are the standard Schwarzschild coordinates.
Let us denote the $r$ value of the particle's location at time $t$ by $r_{\rm p}(t)$.
With suitable boundary/initial conditions, a unique solution is obtained for
$\phi_{lm}$ (for each $l,m$), which is continuous along $r=r_{\rm p}(t)$.
However, the derivatives $\phi_{lm,r}$ and $\phi_{lm,t}$ will generally suffer
a finite jump discontinuity across $r=r_{\rm p}(t)$, which reflects the presence of a
source ``shell'' resulting from decomposing the point charge in spherical
harmonics. In particular, if the orbit is eccentric, the derivatives of
$\phi_{lm}$ will generally be discontinuous functions of $t$ at a fixed value of
$r$ along the orbit.

Now imagine trying to reconstruct $\phi_{lm}(t,r)$ (for some fixed $r$ along
the orbit) as a sum over its Fourier components:
\begin{equation} \label {Eq:12}
\phi_{lm}(t,r)=\sum_{\omega} R_{lm\omega}(r)e^{-i\omega t}.
\end{equation}
Since, for an eccentric orbit, $\phi_{lm}(t,r)$ is only a $C^0$ function of $t$
at the particle's worldline, it follows from standard Fourier theory \cite{James}
that the Fourier sum in Eq.\ (\ref{Eq:12})
will only converge there like $\sim\omega^{-1}$.
The actual situation is even worse, however, because for self-force calculations we need
not only $\phi_{lm}(t,r)$ but also its derivatives.
For instance, to calculate the $r$ component of the self force we need to evaluate
(one-sided limits of) $\phi_{lm,r}(t,r)$ as $r$ approaches $r_{\rm p}(t)$.
Suppose again that we want to reconstruct
this quantity from its Fourier components (namely the functions $R_{lm\omega,r}$).
Since $\phi_{lm,r}(t,r)$ is a discontinuous function of $t$, we will inevitably face here
the well known ``Gibbs phenomenon'' \cite{Gibbs}. Namely, the Fourier sum will
fail to converge to the right value at $r\to r_{\rm p}(t)$.
(The problematic behavior of the Fourier sum is simply a consequence of
our attempt to construct a discontinuous function---or a non-smooth function in the
case of the field $\phi_{lm}$ itself---from a sum over smooth harmonics.)

From a practical point of view this would mean that
(i) at the coincidence limit $r\to r_{\rm p}(t)$
the sum over $\omega$ modes would fail to yield the correct one-sided values of
$\phi_{lm,r}(t,r)$, however many $\omega$ modes are included in the sum; and
(ii) if we reconstruct $\phi_{lm,r}$ at a point $r=r_{0}$ off the worldline,
then the Fourier series should indeed converge; Alas, the number of $\omega$ modes
required for achieving a prescribed precision would grow unboundedly as $r_{0}$
approaches $ r_{\rm p}(t)$, making it extremely difficult to evaluate $\phi_{lm,r}$
at the coincidence limit.

This technical difficulty is rather generic, and will show also in calculations of the
local EM and gravitational fields.
Consider, as a second example, the gravitational perturbation from a point mass
moving in an eccentric orbit in Schwarzschild:
In suitable gauges (like the Lorenz gauge, often applied in self-force
calculations) the multipole $lm$-modes of the physical metric perturbation
{\footnote{Formally, the relevant multipoles in this case are not the standard
spherical harmonics but rather the 2nd-rank tensorial harmonics. Similarly, in
the discussion below regarding Teukolsky fields in Schwarzschild, the relevant
multipoles are the spin-weighted spherical harmonics. These technical details
do not in any way affect the discussion here.}}
are again $C^0$ functions of $r$ and $t$ along the orbit, and their derivatives are generally
discontinuous there. Attempting to construct them naively from a sum over frequency modes
would encounter the same difficulty as in the scalar case:
A poor convergence of the ($lm$-modes of the) metric perturbations,
and lack of convergence for their derivatives.

For orbits in Kerr spacetime the situation is basically similar though more subtle.
The Kerr variant of the mode-sum method \cite{Barack:2002mh}
requires, just as in its Schwarzschild counterpart, the {\it spherical}
harmonic modes $\phi_{lm}(r,t)$ of the perturbation field (as well as their
derivatives) as input.{\footnote {Although the full separation of the field
equation in Kerr is based on the spheroidal harmonics,
the mode sum method (at least in its present form \cite{Barack:2002mh})
is based on the spherical harmonics.
Note that the latter harmonics do provide a valid mode decomposition even in Kerr,
even though the field equation couples modes of different spherical-harmonic $l$.
One of the possible practical ways to construct a Fourier-spherical-harmonic mode
$\phi_{lm\omega}$ is by solving for the separable spheroidal-harmonic modes $l'm\omega$
for various $l'$, and then summing over their contributions to
$\phi_{lm\omega}$, as discussed in Ref.\ \cite{Barack:2002mh}.
}}
For given $l m$, the spherical-harmonic decomposition of the point charge will
again result in a $\delta$-function-type source term distributed over a shell,
which in turn renders the derivatives of $\phi_{lm}(r,t)$ discontinuous.
Therefore an attempt to construct $\phi_{lm}(r,t)$ (and, more crucially, its derivatives)
through a naive summation over its Fourier modes
will lead to the same difficulties as in the Schwarzschild case.

The problem discussed here takes an even more extreme form when considering EM
or gravitational perturbations via the Teukolsky formalism:
Here, the $lm$ modes of the perturbation fields (now the Newman-Penrose fields
$\varphi_0,\varphi_2$ or $\Psi_0,\Psi_4$) are not even continuous at the particle's
orbit---a consequence of the fact that the source term for Teukolsky's equation
involves derivatives of the electric four-current or the energy-momentum tensor
associated with the particle (a single derivative in the EM case; a second
derivative in the gravitational case).{\footnote{Note also that the gravitational
self-force calculation requires the local metric perturbation, and the construction
of the latter requires one to apply certain differential operators to the
Teukolsky fields $\Psi_0$ or $\Psi_4$ \cite{nowski}. The situation with the
EM self-force is similar, as it again requires applying certain differential operators
to $\varphi_0$ or $\varphi_2$. }}
Again, a naive attempt to construct these multipoles as a sum over their $\omega$ modes
will be hampered by the Gibbs phenomenon, and the associated lack of convergence.

In this article we propose a way around the above problem, which is both elegant
and extremely simple. In our method we use the {\it homogeneous} radial functions
$R_{lm\omega}(r)$ (extended all the way through to the particle's worldline),
instead of the actual inhomogeneous functions. The Fourier sum of these homogeneous
radial functions is found to converge exponentially-fast, and to yield the
correct values of the perturbation multipoles (and their derivatives) along
the particle's worldline. We shall focus in this paper on the scalar-field case.
We justify our new method using simple analytical arguments, and then
demonstrate the validity of the method (and the exponential convergence)
with an explicit numerical calculation in the case $l=0$.
The same method should be applicable, however, for
any of the other problems mentioned above: EM and gravitational perturbations
using Teukolsky's equation (or Sasaki--Nakamura's equation), as well as metric
perturbations in the Lorenz gauge. A forthcoming paper \cite{BGSprep} will report on the
computation of the local monopole and dipole modes of the Lorenz-gauge perturbation
(for eccentric orbits in Schwarzschild), facilitated by the new method suggested here.

This paper is structured as follows. In Sec.\ II we set up the physical
scenario---a pointlike scalar charge in a bound orbit around a Schwarzschild
black hole---and review the formalism commonly used in this case to construct
the scalar-field multipoles and the scalar self-force.
Section III demonstrates how the naive sum over frequencies leads to the
Gibbs phenomenon and to the associated problematic convergence at the particle's
location.
Then in Sec.\ IV we present our new method of extended homogeneous
solutions, and show how it cures the problematic behavior of the Fourier sum.
We provide the theoretical justification to this method, as well as numerical
verification in the monopole ($l=0$) case. In Sec.\ V we highlight
the advantages of the new method and discuss foreseeable applications.
Appendices \ref{AppA}--\ref{AppD} give details of the methods used for
our numerical illustrations, and App.\ \ref{AppC} contains some technical
details relating to the formal justification of our new method.

Throughout this work we use standard geometrized units (with $c=G=1$) and metric
signature $({-}{+}{+}{+})$.

\section{Preliminaries}\label{Sec:Preliminaries}

\subsection{Physical setup and scalar-field equation}

Consider a pointlike particle which moves on an eccentric, bound geodesic
orbit around a Schwarzschild black hole with mass parameter $M$. The particle's
worldline is denoted $x_{\rm p}^{\mu}(\tau)$, where $\tau$ is the proper time.
The particle's trajectory is bounded within the range $r_{\rm min}\leq r
\leq r_{\rm max}$ for certain $r_{\rm max}$ and $r_{\rm min}>4M$.
Without loss of generality we shall take the orbit to be equatorial,
namely $\theta_{\rm p}=\pi/2$.

Assume now that the particle carries a scalar charge $q$. This charge couples to
a massless, minimally-coupled scalar field $\Phi(x^{\mu})$, satisfying the
field equation
\begin{equation} \label {Eq:a5}
\Box \Phi=-4 \pi \rho .
\end{equation}
Here $\rho$ is the scalar charge density, which takes the form of a $\delta$-function
along the particle's worldline:
\begin{equation} \label {Eq:a15}
\rho(x)=q\int_{-\infty}^{\infty}\delta^4[x-x_{\rm p}(\tau)] [-g(x)]^{-1/2} d\tau,
\end{equation}
where $g=-r^{4}\sin^{2}\theta$ is the metric determinant, and hereafter the
vectorial indices of $x^{\mu}(\tau)$ and $x_{\rm p}^{\mu}(\tau)$ are
suppressed for brevity.

Since $t$ is timelike (hence monotonic) we can use it instead of $\tau$ to parametrize
the orbit. In the $r,t$ plane the orbit is then denoted by $r=r_{\rm p}(t)$.
Transforming the integration variable in Eq.\ (\ref{Eq:a15}) from $\tau$ to $t$
and substituting $\theta_{\rm p}=\pi/2$, we find
\begin{equation} \label {Eq:a17}
\rho=q(r^{2} u^{t})^{-1}
\, \delta[r-r_{\rm p}(t)] \, \delta[\varphi-\varphi_{\rm p}(t)]
\, \delta[\theta-\pi/2] ,
\end{equation}
where $u^{t} \equiv dt_{\rm p}/d\tau$. Note that $u^{t}$ only depends on $r_{\rm p}(t)$:
We have $u^{t}=E[1-2M/r_{\rm p}(t)]^{-1}$, where $E$ is a constant of motion.

\subsection{Spherical-harmonics decomposition}

We now separate the field equation (\ref{Eq:a5}) by decomposing $\Phi$ in
spherical harmonics, in the form
\begin{equation} \label {Eq:a10}
\Phi=\sum_{l=0}^{\infty}\sum_{m=-l}^{l}\phi_{lm}(t,r)Y_{lm}(\theta,\varphi)/r.
\end{equation}
Here $Y_{lm}$ are the standard (complex-valued) normalized spherical harmonics
given by
$Y_{lm}=c_{lm} P_{lm}(\cos \theta) e^{im \varphi}$,
where $P_{lm}$ are the associated Legendre polynomials and $c_{lm}$ are (real)
normalization constants. The factor $1/r$ is introduced for later convenience.
The charge density in Eq.\ (\ref{Eq:a17}) is decomposed in a similar manner:
\begin{equation} \label {Eq:a20}
\rho=\sum_{l=0}^{\infty}\sum_{m=-l}^{l}\hat\rho_{lm}(t,r)Y_{lm}(\theta,\varphi),
\end{equation}
where
\begin{equation} \label {Eq:a22}
\hat\rho_{lm}(t,r)=\int \rho \, Y^{*}_{lm} d\Omega
=\hat c_{lm} q (r^{2}u^{t})^{-1} e^{-im \varphi_{\rm p}(t)} \delta[r-r_{\rm p}(t)].
\end{equation}
Here $\hat c_{lm}=c_{lm}P_{lm}(0)$, $d\Omega=\sin\theta\, d\theta d\varphi$
is an element of solid angle, and an asterisk denotes complex conjugation.
The field equation (\ref{Eq:a5}) is now separated, and for each $lm$ the
function $\phi_{lm}(t,r)$ satisfies the partial differential equation
\begin{equation} \label {Eq:a25}
\frac{\partial^{2}\phi_{lm}}{\partial r_{*}^{2}}
-\frac{\partial^{2}\phi_{lm}}{\partial t^{2}}-V_l(r)\, \phi_{lm}
= -4 \pi r f(r) \hat\rho_{lm} ,
\end{equation}
where $f(r)\equiv 1-2M/r$, $r_{*}\equiv r+2M\ln[r/(2M)-1]$ is the tortoise
radial coordinate, and
\begin{equation} \label {Eq:a26}
V_{l}(r)=f(r)\left(\frac{2M}{r^{3}}+\frac{l(l+1)}{r^{2}}\right) .
\end{equation}

The function $\phi_{lm}(t,r)$ is determined for each $lm$ by Eq.\ (\ref{Eq:a25}),
supplemented with suitable boundary conditions at null infinity (no incoming waves)
and at the event horizon (no outgoing waves). Since the source term has a
$\delta$-function form, the function $\phi_{lm}(t,r)$ must be continuous at
$r=r_{\rm p}(t)$, but it will generally fail to be differentiable there:
Its $r$ derivative (and also its $t$ derivative, except at the two orbital
turning points) will suffer a discontinuity along the orbit. On each side of
the worldline $r=r_{\rm p}(t)$, however, the function $\phi_{lm}(t,r)$
satisfies the homogeneous part of Eq.\ (\ref{Eq:a25}), and, since the homogeneous
field equation and the curve $r=r_{\rm p}(t)$ are both analytic, we expect
$\phi_{lm}(t,r)$ to be analytic (in both $r$ and $t$) anywhere off the
worldline.
We shall assume this analyticity here, although we are not aware of a proof.
The alternative option appears highly unlikely,
because it would mean that the actual field produced by the
point charge somehow develops irregularities in the vacuum region off
the particle.

\subsection{Self force via mode sum}

Once the functions $\phi_{lm}$ have been determined (e.g.\ numerically),
the self force acting on the scalar
charge may be constructed by the mode-sum method.
This procedure is described in detail in Refs.\ \cite{Barack:1999wf,Barack:2001gx,
Barack:2002mh}. Here we outline it briefly, in order to provide some perspective
on how the quantities $\phi_{lm}$ are incorporated in the construction of the
self force.

Let $\Phi_{l}$ denote the contribution of an individual $l$ to $\Phi$:
\begin{equation} \label {Eq:a35}
\Phi_{l}(t,r,\theta,\varphi) =\sum_{m=-l}^{l}\phi_{lm}(t,r)Y_{lm}(\theta,\varphi)/r.
\end{equation}
The $l$-component of the {\it full-force} field, $F_{l \mu}(x)$, is then
obtained by applying a certain linear differential operator
${\cal F}_{\mu}$ to $\phi_{l}$. [${\cal F}_{\mu}$ is the same differential
operator which determines the force that a smooth, ``non-self'' field $\Phi(x)$
would exert on a test charge.] In the scalar-field case we have
${\cal F}_{\mu}=q \partial_{\mu}$, and therefore
{\footnote{Strictly speaking, $F_{l \mu}$ has been defined in Ref.\
\cite{Barack:1999wf} to be the $l$-multipole of ${\cal F}_{\mu}\Phi$,
rather than ${\cal F}_{\mu}\Phi_{l}$.
Here, however, the $\theta$ component of the self force vanishes as we consider
an equatorial orbit; and for the three remaining components the two definitions
coincide.}
\begin{equation} \label {Eq:a40}
F_{l \mu}={\cal F}_{\mu}\Phi_{l}
=q \sum_{m=-l}^{l}\partial_{\mu}[\phi_{lm}(t,r)Y_{lm}(\theta,\varphi)/r] .
\end{equation}
The quantities $F_{l \mu}$ are to be evaluated at the particle's location.
To be more specific, let us denote by $x_{f}$ the event (on the particle's worldline)
at which the self force is to be evaluated. Then the quantities $F_{l \mu}$ are to be
evaluated at $x\to x_{f}$ (a somewhat rough statement which will be refined shortly).
From Eq.\ (\ref{Eq:a40}) it is obvious that $F_{l \mu}$ depends linearly on the
following functions of $r$ and $t$ (for each $m$):
$\phi_{lm}$, $\phi_{lm,r}$, and $\phi_{lm,t}$.
We use the symbol $\phi_{lm}^{i}$ ($i=0,1,2$) as an abbreviated notation for
these three key functions.

As was discussed above, the quantities $\phi_{lm,r}$ and $\phi_{lm,t}$
are not truly defined on the curve $r = r_{\rm p}(t)$, and in particular at $x=x_{f}$.
Instead, each of these functions has two well-defined (but generally different)
one-sided limits, corresponding to approaching the worldline point $x_{f}$ from the range
$r > r_{\rm p}(t)$ or $r < r_{\rm p}(t)$.
We denote these two one-sided limits as
$x\to x_{f +}$ and $x\to x_{f -}$, respectively.
Correspondingly, the quantities $F_{l \mu}$ will each have two one-sided limits,
$F_{l \mu}^{+}$ and $F_{l \mu}^{-}$, defined by
\begin{equation} \label {Eq:a42}
F_{l \mu}^{\pm}(x_{f}) \equiv \lim_{x\to x_{f \pm}}F_{l \mu}(x).
\end{equation}
The self force at $x=x_{f}$ may now be derived from either set of quantities,
$F_{l \mu}^{+}$ or $F_{l \mu}^{-}$, via the mode-sum formula \cite{Barack:1999wf}
\begin{equation} \label {Eq:a45}
F^{\rm self}_{\mu}= \sum_{l=0}^{\infty}
\left[ F_{l \mu}^{ \pm }(x_f)\mp(l+1/2)A_{\mu}-B_{\mu} \right] ,
\end{equation}
where $A_{\mu}$ and $B_{\mu}$ are certain parameters (``regularization parameter'')
which Refs.\ \cite{Barack:2001gx,Barack:2002mh} determine analytically.
Note that the two one-sided limits in Eq.\ (\ref{Eq:a45}) yield the same value
of $F^{\rm self}_{\mu}$.

 \subsection{Frequency-domain analysis}

The partial differential equation (\ref{Eq:a25}), which determines the functions
$\phi_{lm}(t,r)$, may be tackled in either the time domain or the frequency domain.
In {\it time-domain} calculations, one directly integrates this equation numerically,
using time-evolution on a two-dimensional grid. In the {\it frequency-domain} method,
on the other hand, one first further separates this equation into Fourier frequency
modes, using
\begin{equation} \label {Eq:a50}
\phi_{lm}(t,r)=\int d\omega \, R_{lm \omega}(r) e^{-i \omega t}
\end{equation}
and
\begin{equation} \label {Eq:a55}
-4 \pi r f(r) \hat\rho_{lm}(t,r)=\int d\omega \, Z_{lm \omega}(r) e^{-i \omega t}.
\end{equation}
Equation (\ref{Eq:a25}) then reduces to the {\it ordinary} differential equation
\begin{equation} \label {Eq:a60}
\frac{d^{2}R_{lm\omega}}{d r_{*}^{2}}
-[V_{l}(r)-\omega^{2}] R_{lm\omega}
=Z_{lm \omega} .
\end{equation}
Since $\hat\rho_{lm}(t,r)$ only has support on the curve $r=r_{\rm p}(t)$,
it follows that $Z_{l m \omega}(r)$ is only supported within the range
$r_{\rm min} \leq r \leq r_{\rm max}$.

From Eq.\ (\ref{Eq:a22}) it is evident that $\hat\rho_{lm}$ only depends on $t$
through $r_{\rm p}(t)$ and $\varphi_{\rm p}(t)$.
For an eccentric geodesic $r_{\rm p}(t)$ is periodic,
and $\varphi_{\rm p}(t)$ also has its inherent $2\pi$ periodicity.
It then follows (see App.\ \ref{AppC}) that $\hat\rho_{lm}$ is 2-periodic in $t$,
namely it has a discrete spectrum of the form
$\omega= n\Omega_r+m\Omega_\varphi \equiv \omega_{nm}$.
Here $\Omega_r$ and $\Omega_\varphi$ are the two fundamental frequencies
associated with the particle's radial and azimuthal motions, respectively.
The integrals in Eqs.\ (\ref{Eq:a50}) and (\ref{Eq:a55}) thus reduce to
summation over $n$. In particular,
\begin{equation} \label {Eq:a52}
\phi_{lm}(t,r)=\sum_{n} R_{l m \omega_{nm}}(r) e^{-i \omega_{nm} t},
\end{equation}
where in principle the summation is over all integer values of $n$.

The physically-acceptable solutions of Eq.\ (\ref{Eq:a60}) are those satisfying
the appropriate boundary conditions at both edges $r \to \infty$ and $r \to 2M$,
which correspond to pure outgoing waves at spatial infinity and pure incoming
waves at the event horizon. The standard procedure for constructing the desired
physical solution, for given $lm\omega$, begins with
the construction of a basis of two independent homogeneous solutions,
$R_{lm\omega}^+$ and $R_{lm\omega}^-$.
These two homogeneous solutions satisfy the required boundary conditions at,
respectively, $r \to \infty$ and $r \to 2M$
(note that there is no non-trivial homogeneous solution which satisfies
the required boundary conditions at both edges).
One then utilizes the standard Wronskian-based formula for generating
inhomogeneous solutions to second-order linear differential equations.
Transforming the integration variable from $r_{*}$ to $r$ using
$dr/dr_{*}=f(r)$,
and recalling the bounded support of $Z_{l m \omega}(r)$, this formula takes the form
\begin{eqnarray} \label {Eq:a70}
R_{lm\omega}(r)&=&R_{lm\omega}^+(r)
\int_{r_{\rm min}}^{r} \frac{R_{lm\omega}^-(r') Z_{l m \omega}(r')}
{Wf(r')}\, dr'
+
R_{lm\omega}^-(r)
\int_{r}^{r_{\rm max}} \frac{R_{lm\omega}^+(r') Z_{l m \omega}(r')}
{Wf(r')}
\, dr'
\nonumber\\
&\equiv& R_{lm\omega}^{\rm inh}(r) ,
\end{eqnarray}
where
\begin{equation} \label {Eq:aw}
W\equiv R_{lm\omega}^- \,(dR_{lm\omega}^+/d r_{*})
- R_{lm\omega}^+\, (dR_{lm\omega}^-/dr_{*})={\rm const}
\end{equation}
is the Wronskian. 
In the regions  $r \leq r_{\rm min}$ and $r \geq r_{\rm max}$
this formula reduces to the homogeneous solutions
\begin{equation} \label {Eq:a75}
R_{lm\omega}(r)=\left\{
\begin{array}{ll}
C_{lm\omega}^- R_{lm\omega}^-(r) \equiv {\tilde R}_{lm\omega}^-(r)  ,
& \quad r\leq r_{\rm min}, \\
\\
C_{lm\omega}^+ R_{lm\omega}^+(r) \equiv {\tilde R}_{lm\omega}^+(r)   ,
& \quad r\geq r_{\rm max},
\end{array}
\right.
\end{equation}
where the coefficients $C_{lm\omega}^-$ and $C_{lm\omega}^+$ are given by
\begin{equation} \label {Eq:a80}
C_{lm\omega}^{\pm}=W^{-1}
\int_{r_{\rm min}}^{r_{\rm max}} \frac{R_{lm\omega}^{\mp}(r) Z_{lm\omega}(r)}
{f(r)}
\, dr\, .
\end{equation}

We conclude this section by explicitly writing the frequency-domain expressions for
the three key functions $\phi_{lm}^{i}$, in the particle's neighborhood:
\begin{equation} \label {Eq:a90}
\phi_{lm}(t,r)=\sum_{n} R^{\rm inh}_{l m \omega_{nm}}(r)\,
e^{-i \omega_{nm} t},
\end{equation}
\begin{equation} \label {Eq:a95}
\phi_{lm,r}(t,r)=\sum_{n}\frac{d}{dr} R^{\rm inh}_{l m \omega_{nm}}(r)\,
e^{-i \omega_{nm} t},
\end{equation}
\begin{equation} \label {Eq:a100}
\phi_{lm,t}(t,r)=-i \sum_{n}\omega_{nm} R^{\rm inh}_{l m \omega_{nm}}(r)\,
e^{-i \omega_{nm} t}.
\end{equation}
Since the particle resides in the range $r_{\rm min}\leq r\leq r_{\rm max}$,
the radial functions $R^{\rm inh}_{lm\omega}(r)$ involved in these expressions
are truly inhomogeneous [unlike the functions ${\tilde R}_{lm\omega}^{\pm}(r)$
defined in Eq.\ (\ref{Eq:a75}), which are homogeneous].\footnote {Equation
(\ref{Eq:a95}) should be viewed here as the Fourier decomposition of
$\phi_{lm,r}(t,r)$, rather than the result of a term-by-term differentiation
of Eq.\ (\ref{Eq:a90}). The same applies to Eq.\ (\ref{Eq:a100}).}

\section{The high-frequency problem}
\label{Sec:Problem}

\subsection{Statement of the problem}

The functions $\phi_{lm}^{i}(t,r)$,
whose one-sided values are required for the self-force calculation,
are perfectly (one-sided) smooth, even at the coincidence limit $r\to r_{\rm p}(t)$.
Owing to this smoothness, one may naturally expect that there ought to be a way to
calculate the required one-sided quantities in the frequency domain without
referring to large $\omega$ values. Such a method indeed exists, as we explain
in the next section. However, with a straightforward application of the standard
frequency-domain method, based on Eqs.\ (\ref{Eq:a90})--(\ref{Eq:a100}), one finds
that the Fourier series either fails to converge to the correct values (for $\phi_{lm,r}$
and $\phi_{lm,t}$) or converges very slowly (for $\phi_{lm}$) as $r\to r_{\rm p}(t)$.

To demonstrate this convergence problem we consider first the Fourier sum (\ref{Eq:a95})
for $\phi_{lm,r}$, which is required for calculating $F^{\rm self}_{r}$.
[Essentially the same argument applies to Eq.\ (\ref{Eq:a100}) for $\phi_{lm,t}$.]
Suppose that we attempt to evaluate $\phi_{lm,r}$ at a point $x=x_{f}$ on the worldline,
with coordinates $t=t_{f}$ and $r=r_{f}\equiv r_{\rm p}(t_{f})$.
The values of the radial functions $dR^{\rm inh}_{lm\omega}/dr$ at $r=r_{f}$
are just the Fourier components of the function
\begin{equation} \label {Eq:a102}
\phi_{lm,r}^{f}(t)\equiv \phi_{lm,r}(r=r_{f},t).
\end{equation}
Since $\phi_{lm,r}$ is discontinuous at the worldline, $\phi_{lm,r}^{f}(t)$ is
discontinuous at $t=t_{f}$ [as well as at any other $t$ value for which $r_{\rm p}(t)=r_{f}$].
We therefore encounter here the Gibbs Phenomenon \cite{Gibbs}:
If a function $F(t)$ is discontinuous, its Fourier sum will fail to converge to the
correct value at the discontinuity. (Away from the discontinuity the Fourier sum will
converge, but rather slowly and only conditionally: The $n$-th order term will behave
as $1/n$.)

Consider next the convergence properties of the Fourier sum for $\phi_{lm}$ in
Eq.\ (\ref{Eq:a90}). Defining $\phi_{lm}^{f}(t)\equiv \phi_{lm}(r=r_{f},t)$,
we observe that $\phi_{lm}^{f}(t)$ is continuous, yet its derivative is discontinuous at
$t=t_{f}$. Standard Fourier theory \cite{James} then has it that the Fourier sum
will indeed converge (to the correct value) at $r\to r_{\rm p}(t)$, but this convergence
will be rather slow:
The $n$-th term of the Fourier series is expected to behave as $1/n^{2}$.

\subsection{Numerical illustration: Scalar-field monopole}\label{Subsec:old}

It is instructive to illustrate the above problem with an explicit calculation.
For this, we consider the example of the monopole mode $l=m=0$. The spectrum in
this case becomes simply $w=n\Omega_r$ (with integer $n$), and the Fourier sum
(\ref{Eq:a52}) takes the form
~~~~~~~~~~~~~~~~~~~~~~~~~~~~~~~~~~~~~~~~~~~~~~~~~~~~~~~~~
\begin{equation} \label {Eq:72}
\phi(r,t)= \sum_{n=-\infty}^{\infty}R^{\rm inh}_n(r)\,e^{-in\Omega_r t}.
\end{equation}
We hereafter use the notation
$\phi_{l=m=0}\equiv\phi$, $R_{l=m=0,\omega=n\Omega_r}\equiv R_{n}$, etc.\
to represent the various monopole quantities.
For a given orbit, the inhomogeneous $n$-mode radial functions $R^{\rm inh}_n(r)$
can be computed numerically based on Eq.\ (\ref{Eq:a70}). The relevant numerical
procedure is rather standard, and we relegate its description to App.\ \ref{AppD}.
In the following we present sample results and discuss their significance.

Figures \ref{Fig:1} and \ref{Fig:3} display numerical solutions for the sample
orbital parameters $r_{\rm max}=12.5M$ and $r_{\rm min}=(25/3)M\cong 8.333M$.
(This corresponds to ``semi-latus rectum'' $p=10M$ and ``eccentricity'' $e=0.2$,
both quantities defined in App.\ \ref{AppA}.)
In Fig.\ \ref{Fig:1} we plot the (real-valued) partial sums
\begin{equation} \label {Eq:110}
\phi(r,t;n_{\rm max})\equiv
\sum_{n=-n_{\rm max}}^{n_{\rm max}}R^{\rm inh}_n(r)e^{-in\Omega_n t}
\end{equation}
and
\begin{equation} \label {Eq:111}
\phi_{,r}(r,t;n_{\rm max})\equiv
\sum_{n=-n_{\rm max}}^{n_{\rm max}}\frac{d}{dr}R^{\rm inh}_n(r)e^{-in\Omega_n t}
\end{equation}
as functions of $r$ at the fixed time $t$ when the particle is located at
$r=10M$, for a sample of $n_{\rm max}$ values. For comparison, we also plot
the full monopole
solution (and its $r$ derivative), which we obtain using a time-domain numerical
evolution code similar to that developed by Haas in Ref.\ \cite{Haas:2007kz}.
Evidently (and as expected), the convergence of the $n$-mode sum for both $\phi$
and $\phi_{,r}$ seems very fast at $r<r_{\rm min}$ and $r>r_{\rm max}$, but deteriorates
significantly in the domain $r_{\rm min}<r<r_{\rm max}$, where
``Gibbs waves'' dominate the behavior.
Figure \ref{Fig:3} illustrates the convergence properties of the partial sums
$\phi(n_{\rm max})$ and $\phi_{,r}(n_{\rm max})$ at the very location of the particle
(on the same time slice as in Fig.\ \ref{Fig:1}).
The data on the left panel suggest that the partial sum
for the field $\phi$ converges at the particle as $\sim 1/n_{\rm max}$---in accordance with
theoretical expectation \cite{James}. The results shown in the right panel of Fig.\ \ref{Fig:3}
demonstrate that the partial sum for the derivative $\phi_{,r}$ fails to converge to the correct
(one-sided) values. They also (loosely) suggest that this partial sum in fact converges
to the {\em two-side average} value of $\phi_{,r}$ at the particle. This, indeed,
would again accord with theoretical prediction \cite{James}.

\begin{figure}[Htb]
\input{epsf}
\centerline{\epsfysize 7cm \epsfbox{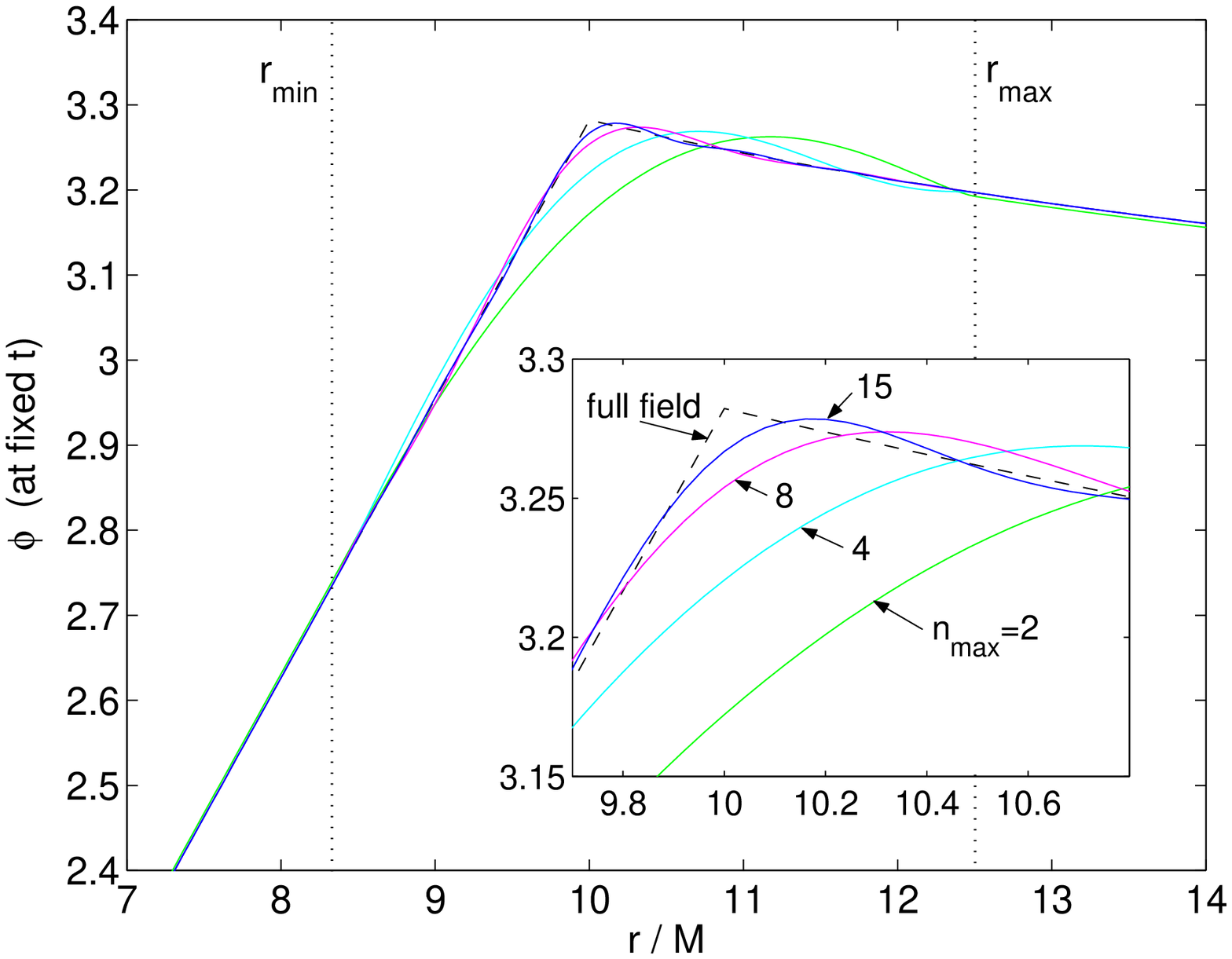}
            \epsfysize 7cm \epsfbox{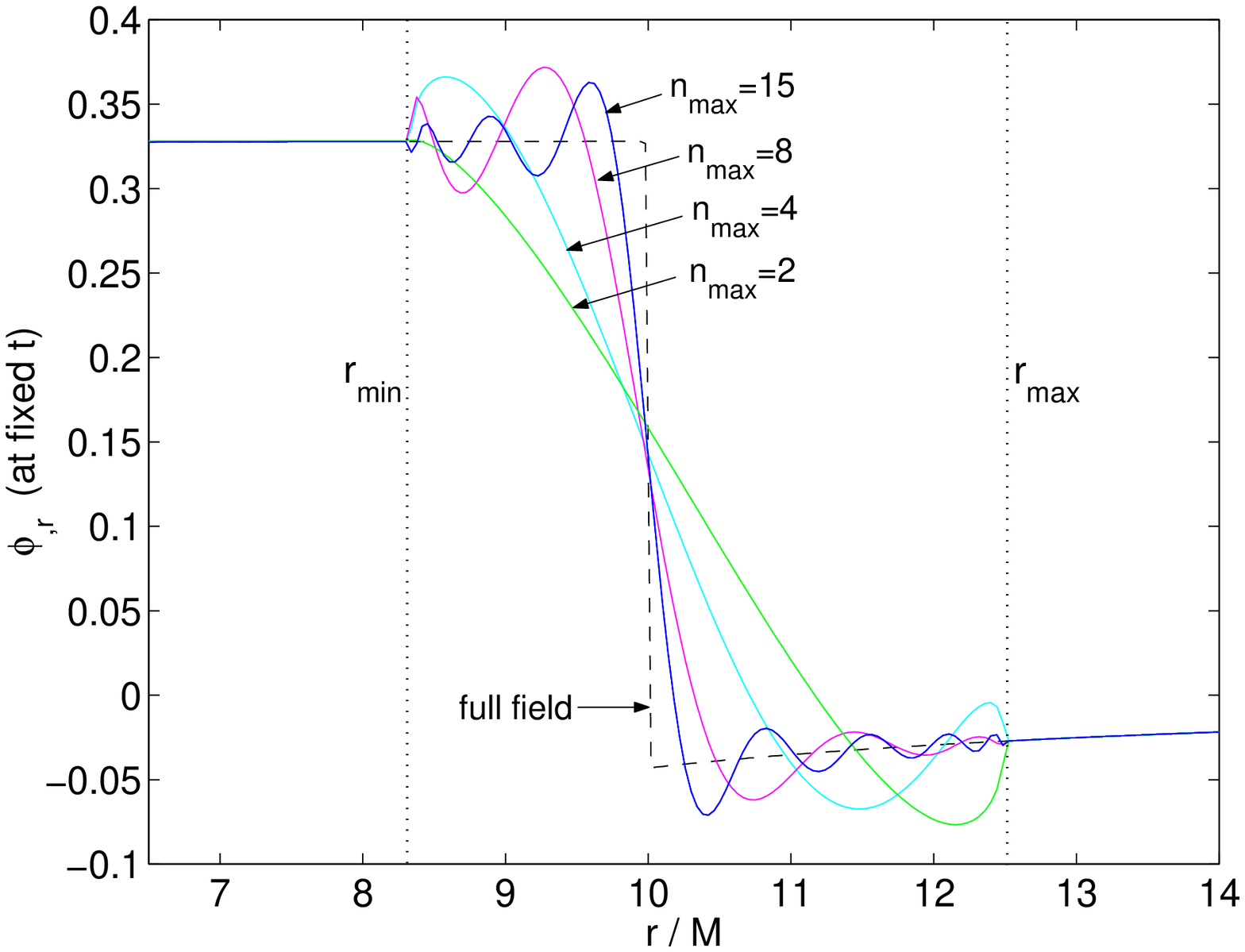}}
\caption{\protect\footnotesize
Construction of the scalar-field monopole and its $r$ derivative as a sum over
inhomogeneous frequency modes in the standard approach. The numerical solutions shown
here correspond to an eccentric geodesic orbit around a Schwarzschild black hole,
with semi-latus rectum $p=10$ and eccentricity $e=0.2$ (see App.\ \ref{AppA}).
The ``periastron'' and ``apastron''
for this orbit are at $r_{\rm min}=(25/3)M\cong 8.333M$ and $r_{\rm max}=12.5M$, respectively.
We used numerical integration to calculate the inhomogeneous radial functions
$R^{\rm inh}_n(r)$ through Eq.\ (\ref{Eq:90}), and then obtained the partial sums
$\phi(t,r;n_{\rm max})$ and $\phi_{,r}(t,r;n_{\rm max})$ as defined in Eqs.\ (\ref{Eq:110})
and (\ref{Eq:111}). Plotted here (solid lines) are the partial sums
(per unit scalar charge) for $n_{\rm max}=2,4,8,15$, as functions of $r$ at a fixed time
$t$ when the particle is at $r=10M$ (this corresponds to radial phase $\chi=\pi/2$;
see App.\ \ref{AppA}).
The {\em left panel} displays the scalar field itself; the right panel shows its
$r$ derivative. For comparison, we also display (dashed line) the {\em full} scalar monopole
solution, which we obtained directly using numerical evolution in the time domain
(for our purpose, this latter solution can be taken as an accurate benchmark).
It is evident that the $n$-mode sum converges quickly to the correct value in the
regions $r<r_{\rm min}$ and $r>r_{\rm max}$, but the convergence deteriorates
inside the domain $r_{\rm min}<r<r_{\rm max}$, where ``Gibbs waves'' set in.
The partial sum over frequency modes is smooth at the particle, and hence, strictly
speaking, cannot recover the true jump discontinuity in the field derivative there.
Finding a way around this technical problem is the main goal of this work.
}
\label{Fig:1}
\end{figure}
\begin{figure}[Htb]
\input{epsf}
\centerline{\epsfysize 7cm \epsfbox{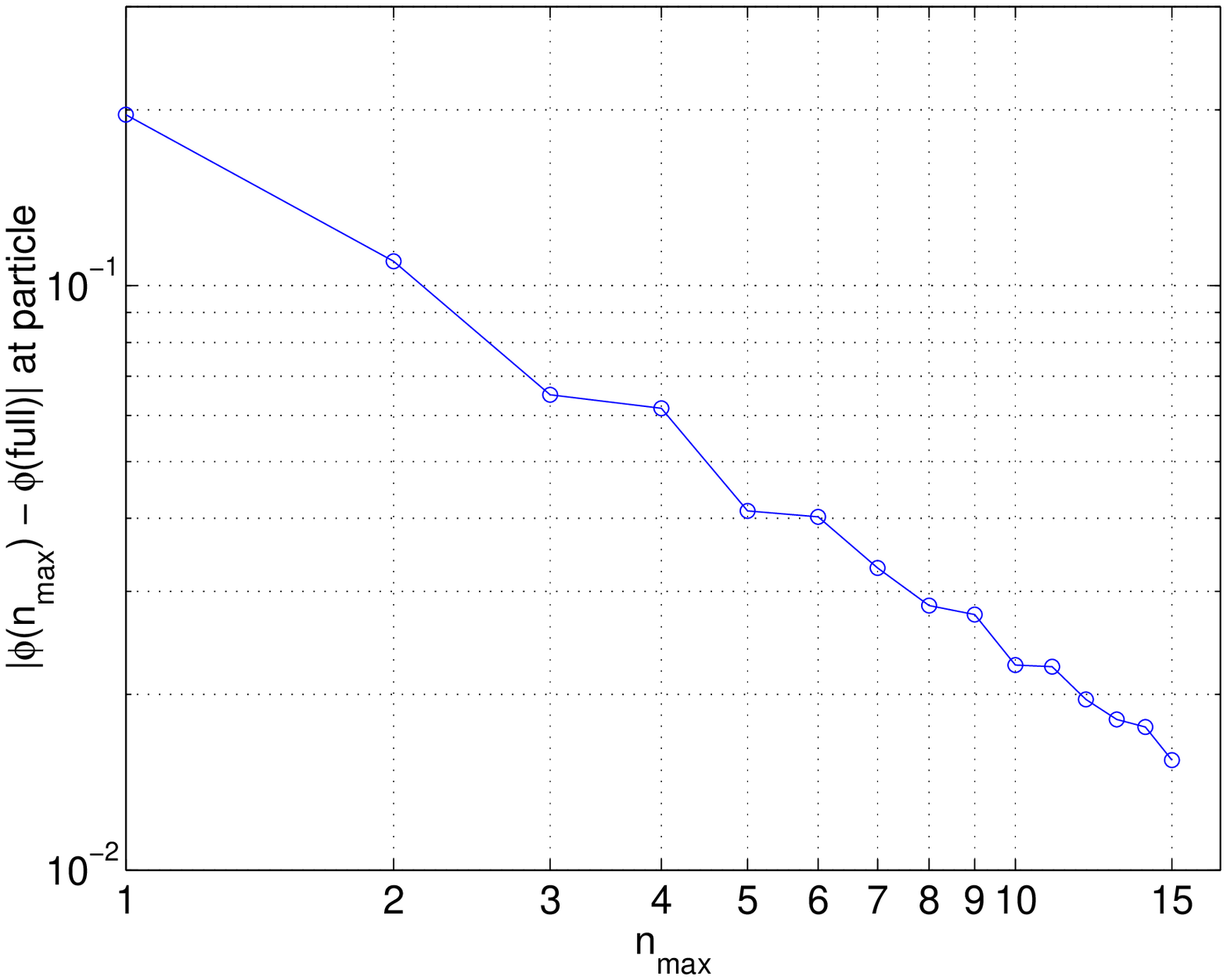}
            \epsfysize 7cm \epsfbox{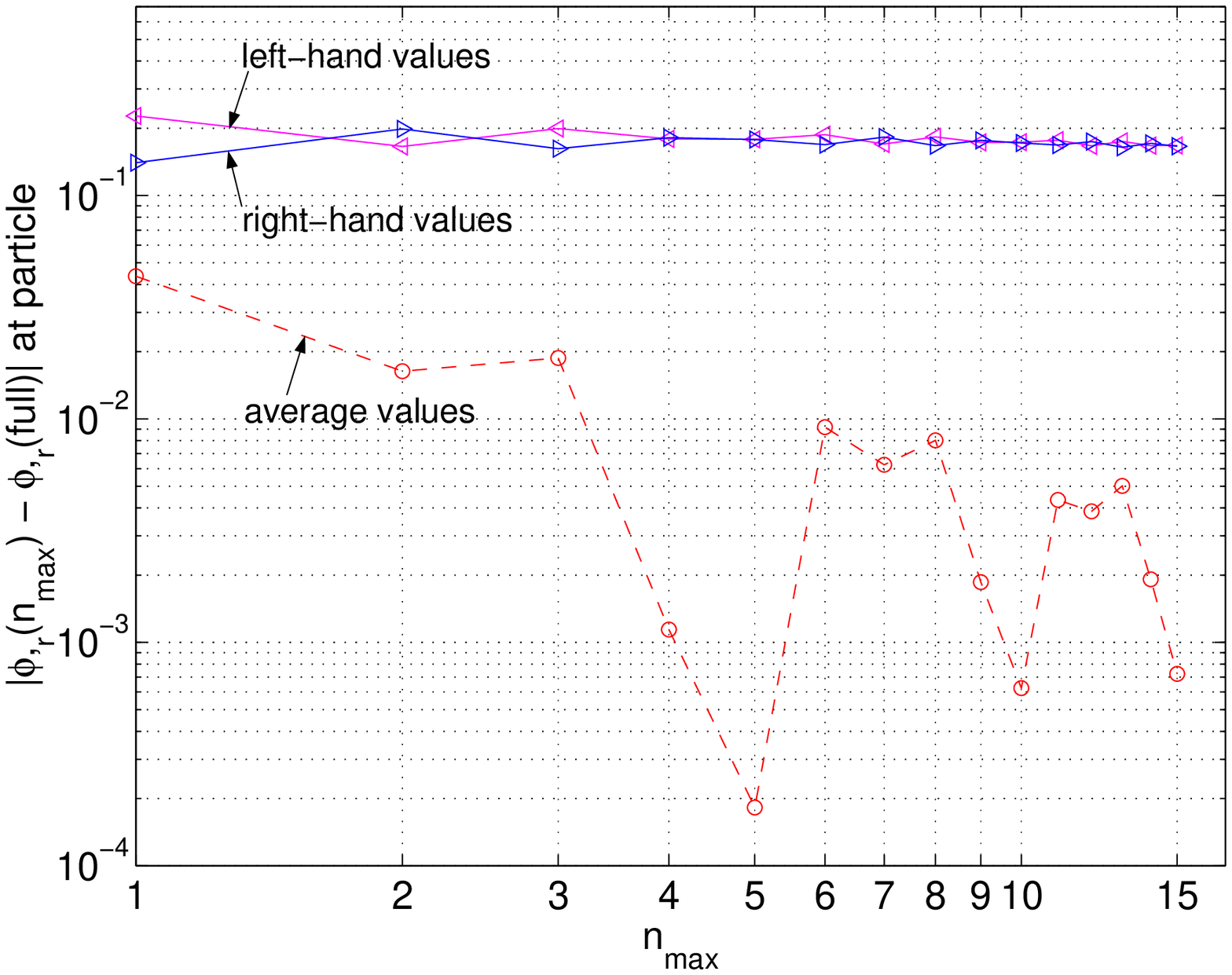}}
\caption{\protect\footnotesize
Convergence of the Fourier $n$-mode sum at the location of the particle
(illustrated here based on the numerical data of Fig.\ {\ref{Fig:1}}).
{\em Left panel:} The deviation (per unit scalar charge) of the partial sum
$\phi(n_{\rm max})$ from the full field $\phi$ {\em at the location of the particle},
as a function of $n_{\rm max}$. The $n$-mode partial sum appears to converge to the
correct value approximately as $1/n_{\max}$ (note the log-log scale), as expected on
theoretical grounds.
{\em Right panel:} The deviation (per unit scalar charge)
of $\phi_{,r}(n_{\rm max})$ from the full-field derivative $\phi_{,r}$,
again evaluated at the particle's location.
Since the true $\phi_{,r}$ [unlike $\phi_{,r}(n_{\rm max})$] has two
different one-sided values at $r=r_{\rm p}$, so does the deviation.
The two solid lines represent these two one-sided values of the deviation.
(These two values actually have
opposite signs for each $n_{\rm max}$, which is obscured here since only the
absolute value of the deviation is shown.)
From the essentially-horizontal shape of the two solid lines it is evident that
the $n$-mode sum for $\phi_{,r}$ does {\em not} converge to the correct one-sided values.
As an aside, we also plot here (dashed line) the difference between
$\phi_{,r}(n_{\rm max})$ and the two-side {\em average} value of the full derivative
$\phi_{,r}$ (per unit scalar charge, at the particle's location).
The graph loosely suggests (referring to its upper envelop and ignoring the
seemingly oscillatory deep structure)
that $\phi_{,r}(n_{\rm max})$ converges, albeit very
slowly, to the average value of $\phi_{,r}$ at the discontinuity. That, indeed,
would again be consistent with theoretical prediction.
}
\label{Fig:3}
\end{figure}

\subsection{Practical implications of the high-frequency problem}

The above numerical example serves to illustrate the following: From a practical point of view,
it would seem very difficult to extract the correct values of the key functions $\phi_{lm}^{i}$
at the particle's location, based plainly on a naive summation over frequency modes as in
Eqs.\ (\ref{Eq:a90})--(\ref{Eq:a100}). The partial Fourier sum for the field $\phi_{lm}$
itself would converge very slowly (as $1/n$) and its evaluation would hence be computationally
expensive. Worse, the partial sums for the derivatives $\phi_{lm,r}$ and $\phi_{lm,t}$ would
simply fail to yield the desired one-sided values at the particle, even if one could sum
over infinitely many modes.

Having stated the above, we should also point out that Eqs.\ (\ref{Eq:a95}) and (\ref{Eq:a100})
should {\em not} be deemed entirely useless for the purpose of calculating $\phi_{lm,r}$ and
$\phi_{lm,t}$ at the particle. In principle, one could pick a point close to $r= r_{\rm p}(t)$,
yet not quite at $r_{\rm p}(t)$, and calculate $\phi_{lm,r}$ (say) there. The Fourier sum will
converge at this point, although rather slowly. One could then pick a series of $r$-values which
approach $r_{\rm p}(t)$ (say, from the `+' side), calculate $\phi_{lm,r}$ at each of these
points, and then evaluate the desired sided-limit of $\phi_{lm,r}$ through extrapolation.
This procedure, however, is cumbersome and is hardly likely to be computationally
tractable.

An alternative implementation strategy could make use of the fact that the sums in Eqs.\
(\ref{Eq:a95}) and (\ref{Eq:a100}) actually converge to the two-side averages of $\phi_{lm,r}$
and $\phi_{lm,t}$ at the particle (recall the discussion relating to Fig.\ \ref{Fig:3}).
These average values (along with $\phi_{lm}$ itself) could be used to construct the average
force modes $\bar F_{l\mu}\equiv (F_{l \mu}^{+}+F_{l \mu}^{-})/2$, which could then
be directly implemented in a ``two-side averaged'' version of the mode sum formula:
$F^{\rm self}_{\mu}= \sum_{l=0}^{\infty}\left[ \bar F_{l \mu}(x_f) - B_{\mu} \right]$.
Although this method is likely to be by far more efficient than the extrapolation
method mentioned earlier, it would still present a computational challenge, as this
method, too, involves the evaluation of slowly-converging Fourier sums.

\section{Method of extended homogeneous solutions}
\label{Sec:Solution}

\subsection{Formulation of method}

In this section we describe our alternative method for frequency-domain construction
of the key quantities $\phi_{lm}^{i}$ required for calculation of the self force.
This method, to which we refer as the method of {\it extended homogeneous solutions},
completely avoids the high-frequency problem described above and ensures exponentially-fast
convergence of the Fourier series.

We begin by extending the definition of the homogeneous functions
${\tilde R}_{lm\omega}^{\pm}(r)$ in Eq.\ (\ref{Eq:a75}) to the entire
domain $r>2M$:
\begin{equation} \label {Eq:Rtilde}
{\tilde R}_{lm\omega}^{\pm}(r)\equiv
C^{\pm}_{lm\omega} R_{lm\omega}^{\pm}(r), \quad\quad r>2M.
\end{equation}
We then define the two {\it time-domain extended homogeneous solutions}
$\tilde\phi_{lm}^{+}$ and $\tilde\phi_{lm}^{-}$ to be
the outcome of replacing $R_{l m \omega}^{\rm inh}$
in Eq.\ (\ref{Eq:a90}) by the homogeneous solutions
${\tilde R}_{lm\omega}^{+}$ or ${\tilde R}_{lm\omega}^{-}$, respectively:
\begin{equation} \label {Eq:a116}
\tilde\phi_{lm}^{\pm}(t,r) \equiv
\sum_{n} {\tilde R}_{lm\omega_{nm}}^{\pm}(r) e^{-i \omega_{nm} t} .
\end{equation}
We emphasize that each of the fields $\tilde\phi_{lm}^{+}$ and $\tilde\phi_{lm}^{-}$
is defined in the entire domain $2M<r<\infty$.

The convergence properties of the sum in Eq.\ (\ref{Eq:a116}) are dictated
by the large-$|n|$ asymptotic behavior of the coefficients ${\tilde
R}_{lm\omega_{nm}}^{\pm}(r)$. This high-frequency asymptotic behavior can
be examined using a WKB-type analysis, which we carry out in App.\ \ref{AppC}.
This analysis shows that, at least within the leading-order WKB approximation,
the terms on the right-hand side of Eq.\ (\ref{Eq:a116}) decay (at least)
{\it exponentially} in $|n|$. This exponential decay is uniform in $t$ and
$r$, throughout $r>2M$. Also, one naturally expects that the contribution from
higher-order terms in the large-$\omega$ WKB expansion will converge
even faster than the leading-order contribution, and hence will not affect
this uniform exponential decay.

The exponential convergence of the sum (\ref{Eq:a116}) is extremely
convenient for numerical applications, as illustrated in the next subsection.
But it also has important mathematical consequences. Since the homogeneous
radial functions ${\tilde R}_{lm\omega}^{\pm}(r)$ are analytic,
the uniform exponential decay of the individual terms in the above sum
implies that the overall sum---namely the extended homogeneous solution
$\tilde\phi_{lm}^{\pm}(t,r)$---is an {\it analytic} function of $r$ and $t$
throughout $r>2M$.

We now argue that
on each side of the curve $r=r_{\rm p}(t)$ the actual time-domain function
$\phi_{lm}(t,r)$ coincides with one of these extended homogeneous solutions, namely
\begin{equation} \label {Eq:a117}
\phi_{lm}(t,r)=\left\{
\begin{array}{ll}
\tilde\phi_{lm}^{+}(t,r)  ,
&\quad r\geq r_{\rm p}(t), \\
\\
\tilde\phi_{lm}^{-}(t,r)  ,
&\quad r\leq r_{\rm p}(t).
\end{array}
\right.
\end{equation}
For concreteness, let us present our argument explicitly referring to the first
of these equalities:
(i)  In the domain $r>r_{\rm max}$ this equality obviously holds
because $R_{lm\omega}^{inh}$ and ${\tilde R}_{lm\omega}^{+}$ coincide in that domain.
(ii)  As was already mentioned in Sec.\ \ref{Sec:Preliminaries}, we assume that the
function $\phi_{lm}(t,r)$ is analytic throughout the range $r>r_{\rm p}(t)$
[as well as in the other range, $2M<r<r_{\rm p}(t)$],
because the alternative option appears unreasonable.
(iii) As was just discussed above, the high-frequency analysis in App.\ \ref{AppC}
strongly suggests that the extended homogeneous functions $\tilde\phi_{lm}^{+}(t,r)$
are analytic throughout $r>2M$.
(iv)  Since both functions $\phi_{lm}$ and $\tilde\phi_{lm}^{+}$ are analytic throughout
the domain $r>r_{\rm p}(t)$ [from (ii) and (iii)], and coincide at $r>r_{\rm max}$ [from (i)],
they must coincide throughout $r>r_{\rm p}(t)$.
(v) By continuity of both functions $\phi_{lm}$ and $\tilde\phi_{lm}^{+}$, they coincide
at $r=r_{\rm p}(t)$ as well.
Obviously, the same line of argument applies to the second
of the equalities (\ref{Eq:a117}) as well.

In the rest of this subsection we describe the utility of the extended homogeneous
fields defined above in calculations of the self-force via the mode-sum method.
Recall from Eqs.\ (\ref{Eq:a40})--(\ref{Eq:a45}) that this method requires as input
(either of) the one-sided limits of $\phi_{lm}^{i}$ at the particle, which we now denote
\begin{equation} \label {Eq:a104}
\phi_{lm}^{i \pm}(x_{f}) \equiv \lim_{x\to x_{f \pm}} \phi_{lm}^{i}.
\end{equation}
In terms of $\phi_{lm}^{i \pm}$, the various components of the quantities
$F_{l \mu}^{\pm}(x_f)$ [as defined through Eqs.\ (\ref{Eq:a40}) and (\ref{Eq:a42})
and used in the mode-sum formula (\ref{Eq:a45})] are expressed directly as
\begin{equation} \label {Eq:a105}
\left\{F_{l t}^{\pm},F_{l r}^{\pm},F_{l \varphi}^{\pm}\right\}
=\frac{q}{r_f} \sum_{m=-l}^{l}
\left\{\phi_{lm,t}^{\pm}, \ \phi_{lm,r}^{\pm}-\phi_{lm}^{\pm}/r_f,
\ i \,m\,\phi_{lm}^{\pm}\right\}\,  Y_{lm}(\pi/2,\varphi_{f})
\end{equation}
(along with $F_{l \theta}^{\pm}=0$).

For concreteness, let us focus first on one of the quantities $\phi_{lm}^{i\pm}$,
say $\phi_{lm}^{+}$.
By definition, the limit $x\to x_{f +}$ in Eq.\ (\ref{Eq:a104})
only samples the range $r>r_{\rm p}(t)$.
Using Eqs. (\ref{Eq:a117}) and (\ref{Eq:a116}) we may re-express $\phi_{lm}^{+}$ as
\begin{equation} \label {Eq:a120}
\phi_{lm}^{+}(x_{f})=\lim_{x\to x_{f +}} \sum_{n}
{\tilde R}_{lm\omega_{nm}}^{+}(r) e^{-i \omega_{nm} t}.
\end{equation}
Since the sum over $n$ here converges uniformly, we may
interchange the limit and summation. However, the functions
${\tilde R}_{lm\omega_{nm}}^{+}$ and $e^{-i \omega_{nm} t}$
are analytic, so we can now omit the limit $x\to x_{f +}$ and instead simply evaluate
these functions at $x=x_{f}$. The final outcome from these manipulations is stated
in (the `+' case of) Eq.\ (\ref{Eq:a125}) below.

The above treatment is equally applicable to $\phi_{lm}^{-}$,
and we obtain a similar formula for constructing $\phi_{lm}^{-}(x_{f})$ out
of the extended homogeneous modes ${\tilde R}_{lm\omega_{nm}}^{-}(r)$
[the `$-$' case of Eq.\ (\ref{Eq:a125})].
Moreover, the same treatment also applies to $\phi_{lm,r}^{\pm}$ and
$\phi_{lm,t}^{\pm}$. The six key quantities $\phi_{lm}^{i \pm}$ can all be
constructed from the extended homogeneous radial functions (and their derivatives)
in the form
\begin{equation} \label {Eq:a125}
\phi_{lm}^{\pm}(x_{f})= \sum_{n}
{\tilde R}_{lm\omega_{nm}}^{\pm}(r_{f}) e^{-i \omega_{nm} t_{f}},
\end{equation}
\begin{equation} \label {Eq:a130}
\phi_{lm,r}^{\pm}(x_{f})= \sum_{n}
\frac{d}{dr}{\tilde R}_{lm\omega_{nm}}^{\pm}(r_{f}) e^{-i \omega_{nm} t_{f}},
\end{equation}
\begin{equation} \label {Eq:a135}
\phi_{lm,t}^{\pm}(x_{f})=- i\sum_{n} \omega_{nm}
{\tilde R}_{lm\omega_{nm}}^{\pm}(r_{f}) e^{-i \omega_{nm} t_{f}}.
\end{equation}

Equations (\ref{Eq:Rtilde}) and (\ref{Eq:a125})--(\ref{Eq:a135}), combined with Eqs.\
(\ref{Eq:a105}) and (\ref{Eq:a45}), constitute our new method of calculating the
self force in the frequency domain. The high-frequency problem is entirely
circumvented in this method, as the Fourier sum converges
exponentially-fast for all functions $\phi_{lm}^{i \pm}$.

\subsection{Numerical illustration: Scalar-field monopole revisited}\label{Subsec:new}

Let us revisit the calculation of the scalar-field monopole---this time using
the method of extended homogeneous solutions. The homogeneous basis solutions
$R^{\pm}_n(r)$ are constructed numerically in just the same manner as in the standard
approach (see App.\ \ref{AppB}). Then, however, instead of calculating the
actual inhomogeneous modes $R^{\rm inh}_n(r)$ as in Sec.\ \ref{Sec:Problem},
we construct the extended homogeneous solutions $\tilde R^{\pm}_n(r)$ as they
are defined in Eq.\ (\ref{Eq:Rtilde}), with the coefficients
$C^{\pm}_{l=m=0,\omega=n\Omega_r}\equiv C^{\pm}_n$ calculated through
Eq.\ (\ref{Eq:C20}) of App.\ \ref{AppD}.
The time-domain extended fields and their $r$ derivatives are then approximated
by the (real-valued) partial sums
\begin{equation} \label {Eq:560}
\tilde\phi^{\pm}(r,t;n_{\rm max})=\sum_{n=-n_{\rm max}}^{n_{\rm max}}
\tilde R^{\pm}_{n}(r)\, e^{-in\Omega_n t},
\end{equation}
\begin{equation} \label {Eq:565}
\tilde\phi_{,r}^{\pm}(r,t;n_{\rm max})=\sum_{n=-n_{\rm max}}^{n_{\rm max}}
\frac{d}{dr}\tilde R^{\pm}_{n}(r)\, e^{-in\Omega_n t},
\end{equation}
with sufficiently large $n_{\rm max}$.

We point out the following matters relating to the implementation of
Eqs.\ (\ref{Eq:560}) and (\ref{Eq:565}): (i) In the new approach, the computation
of $\tilde\phi$ (or $\tilde\phi_{,r}$) for all $r$ and $t$ involves the (numerical)
evaluation of only two integrals---the ones in Eq.\ (\ref{Eq:C20}) of App.\ \ref{AppD};
In contrast, the standard approach requires the evaluation of two integrals---the
ones in Eq.\ (\ref{Eq:90})---{\em
separately for each value of} $r$ between $r_{\rm min}$ and $r_{\rm max}$.
(ii) The full scalar monopole is continuous at $r=r_{\rm p}(t)$;
however, the contributions to the extended functions $\tilde\phi^+$ and
$\tilde\phi^-$ from each individual $n$ mode do {\em not} match continuously
along this curve. Consequently,
for any finite  $n_{\rm max}$, the partial sums for these extended functions
do not match at $r=r_{\rm p}(t)$.
The amplitude of this mismatch is expected to decrease rapidly with growing
$n_{\rm max}$, as the results below indeed demonstrate.

Figures \ref{Fig:4}--\ref{Fig:6} display numerical solutions obtained based
on Eqs.\ (\ref{Eq:560}) and (\ref{Eq:565}). Our goal here is to assess the performance
of the new method against the standard method, and to this end we have chosen for
our numerical experiment the same orbital parameters as in Figs.\ \ref{Fig:1} and
\ref{Fig:3} of Sec.\ \ref{Sec:Problem}. For clarity, we only show the
`$+$' and `$-$' fields in their respective relevant domains, i.e.,
$r\geq r_{\rm p}(t)$ for the former and  $r\leq r_{\rm p}(t)$ for the latter.

Our numerical illustration serves to demonstrate the following:
(i) The sum over `$+$' and `$-$' extended $n$-modes converges quickly to the
correct, full solution everywhere in the respective domains $r\geq r_{\rm p}(t)$
and $r\leq r_{\rm p}(t)$.
(ii) In particular, the mismatch between the values of the `$+$' and `$-$'
partial sums at the particle's location quickly converges to zero with growing
$n_{\rm max}$.
(iii) The convergence of the extended $n$-mode sum is exponential
everywhere---even in the region $r_{\rm min}\leq r\leq r_{\rm max}$;
in particular, it is exponential at the very location of the particle.
This applies to both the field and its derivatives.
(iv) The new scheme completely circumvents the Gibbs effect which
disrupts the convergence of the inhomogeneous $n$-modes in the
standard approach.

\begin{figure}[Htb]
\input{epsf}
\centerline{\epsfysize 7cm \epsfbox{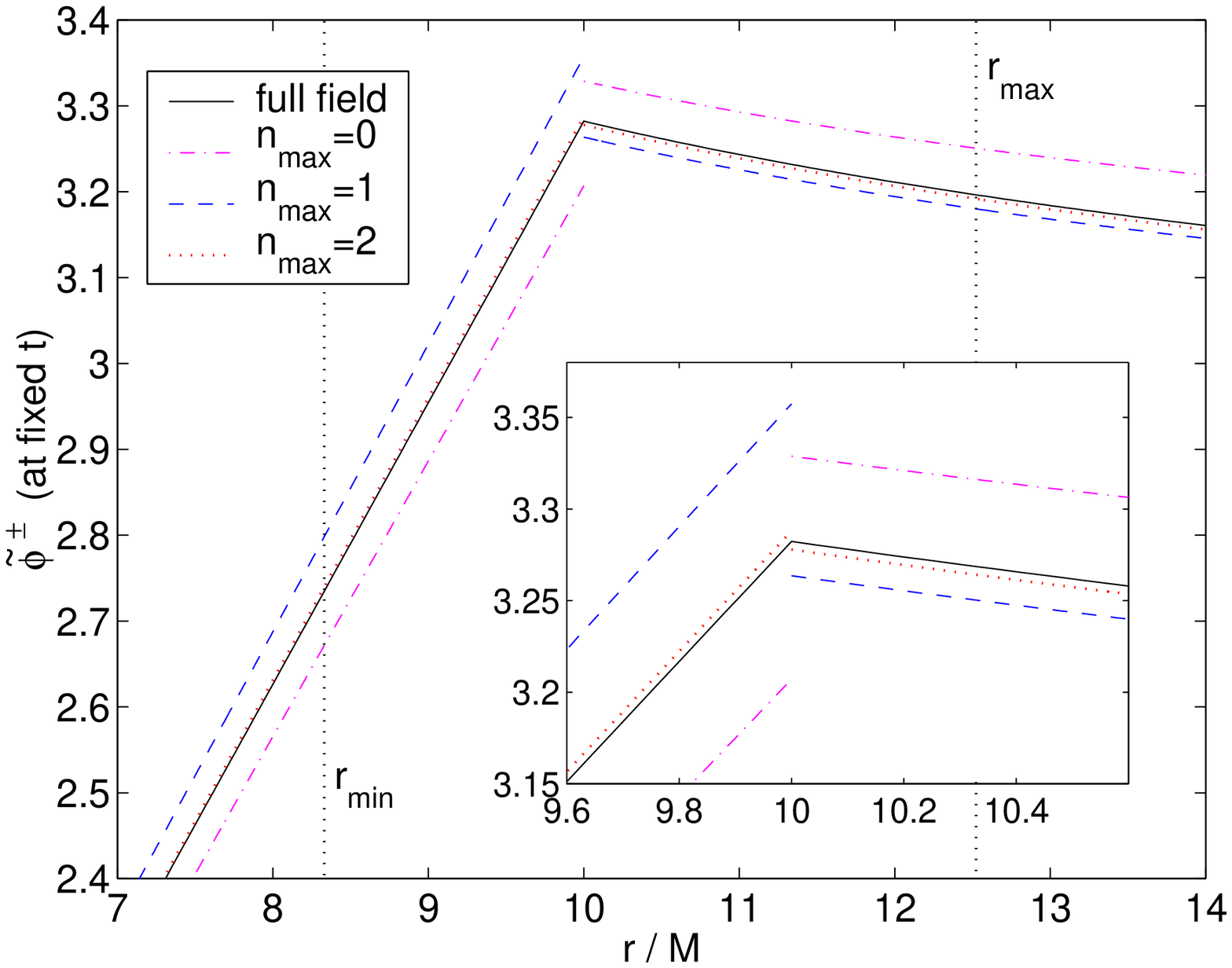}
            \epsfysize 7cm \epsfbox{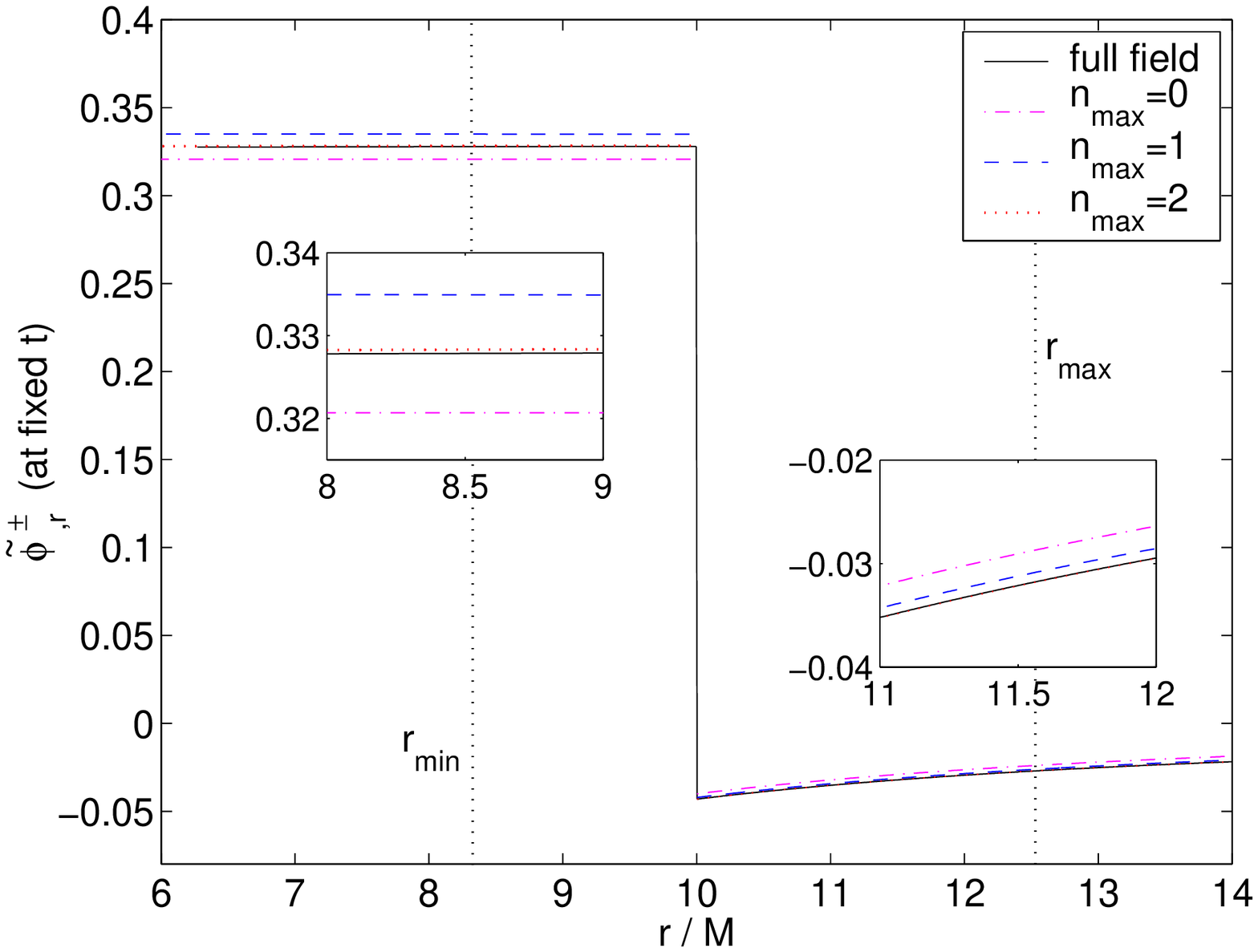}}
\caption{\protect\footnotesize
(To be compared with Fig.\ \ref{Fig:1}.)
Construction of the scalar-field monopole (left panel) and its $r$ derivative
(right panel) using ``extended homogeneous solutions''.
The orbital parameters are chosen here as in Fig.\ \ref{Fig:1},
i.e., $p=10$ and $e=0.2$, and we again present the various fields as functions of
$r$ at a fixed time $t$ when the particle is at $r=10M$.
Vertical dotted lines mark the particle's periastron and apastron radii,
$r_{\rm min}=(25/3)M$ and $r_{\rm max}=12.5M$, respectively.
The solid line represents the full scalar
monopole solution, obtained using numerical evolution in the time domain.
The broken lines represent partial sums over extended homogeneous $n$-modes,
calculated numerically based on Eqs.\ (\ref{Eq:560}) and (\ref{Eq:565}).
For clarity, in both panels we plot the `$+$' and `$-$' partial sums
only in their relevant domains, $r\geq 10M$ and $r\leq 10M$, respectively.
We show here the partial sums
for $n_{\rm max}=0,1,2$ only---the partial sums $\tilde\phi^{\pm}(n_{\rm max}=3)$
are already indistinguishable from the full solution at the scale of this plot
(but see Fig.\ \ref{Fig:5} below). The individual $n$-modes of the extended fields
$\tilde\phi^+$
and $\tilde\phi^-$ do not match continuously at the location of the particle, but
their sum seems to converge quickly, everywhere, to the true solution (which is
continuous). Similar fast convergence is manifest also for the derivative $\phi_{,r}$.
``Gibbs waves'', which disrupt the convergence of the actual inhomogeneous $n$-modes,
are altogether avoided within the new scheme.
}
\label{Fig:4}
\end{figure}
\begin{figure}[Htb]
\input{epsf}
\centerline{\epsfysize 7cm \epsfbox{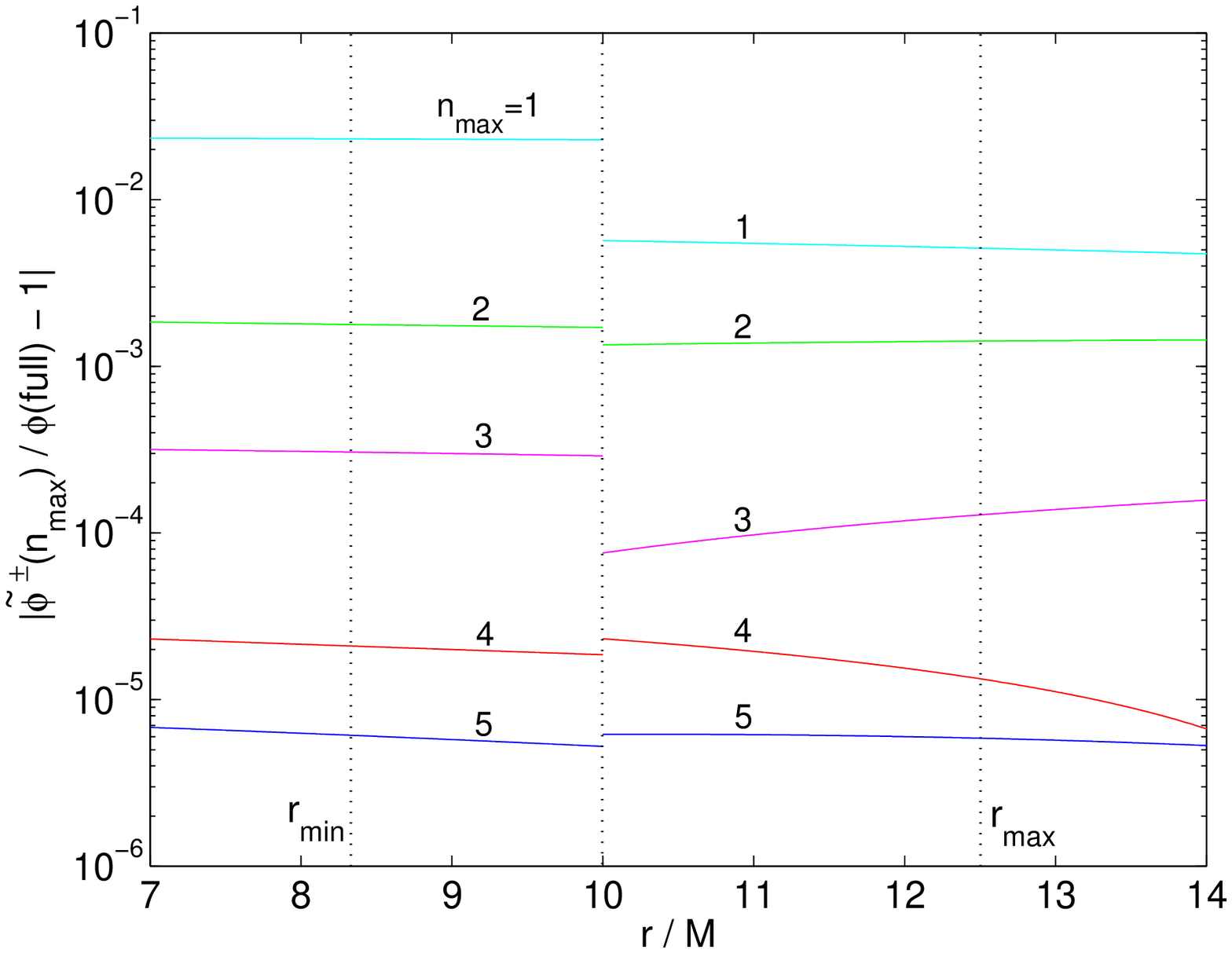}
            \epsfysize 7cm \epsfbox{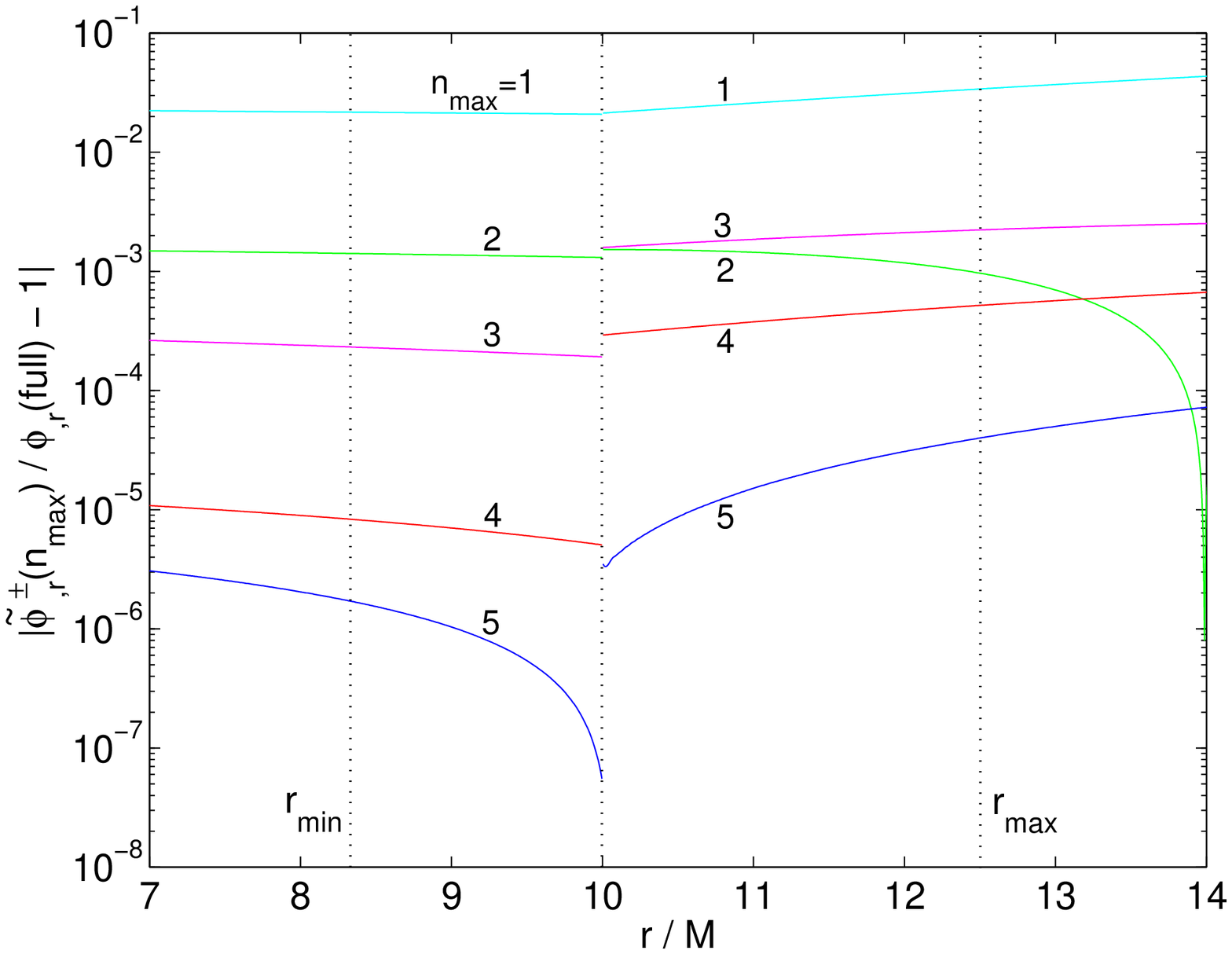}}
\caption{\protect\footnotesize
Convergence of the extended $n$-mode sum for the fields $\tilde\phi^{\pm}$ (left panel)
and their derivatives $\tilde\phi^{\pm}_{,r}$ (right panel). For the same case shown in
Fig.\ \ref{Fig:4}, we plot here the fractional differences between the partial sums
and the full field (or the full-field derivative), for $n_{\rm max}=1$--$5$.
The various graphs are labeled by their corresponding $n_{\rm max}$ values.
The middle vertical dotted line marks the particle's momentary radius at
$r=10M$. Once again, we display the `$+$' and `$-$' values only in
their respective relevant domains $r\geq 10M$ or $r\leq 10M$.
Note the exponential scale of the y-axis.
(The seemingly odd behavior of the data for $n_{\rm max}=2$ and $n_{\rm max}=5$
in the right panel is simply due to a change-of-sign which the corresponding
fractional differences happen to experience around $r=14M$ and $r=10M$, respectively.
The tiny wiggly feature, barely visible near $r=10M$ for $n_{\rm max}=5$, is due
to the numerical error in the time-domain data, which for $\phi_{,r}$ is estimated
at $\sim 10^{-6}$ in fractional terms.) The exponentially-fast convergence
of the extended $n$-mode sum even at $r_{\rm min}\leq r\leq r_{\rm max}$---and
particularly at the very location of the particle---is evident from these plots.
}
\label{Fig:5}
\end{figure}
\begin{figure}[Htb]
\input{epsf}
\centerline{\epsfysize 7cm \epsfbox{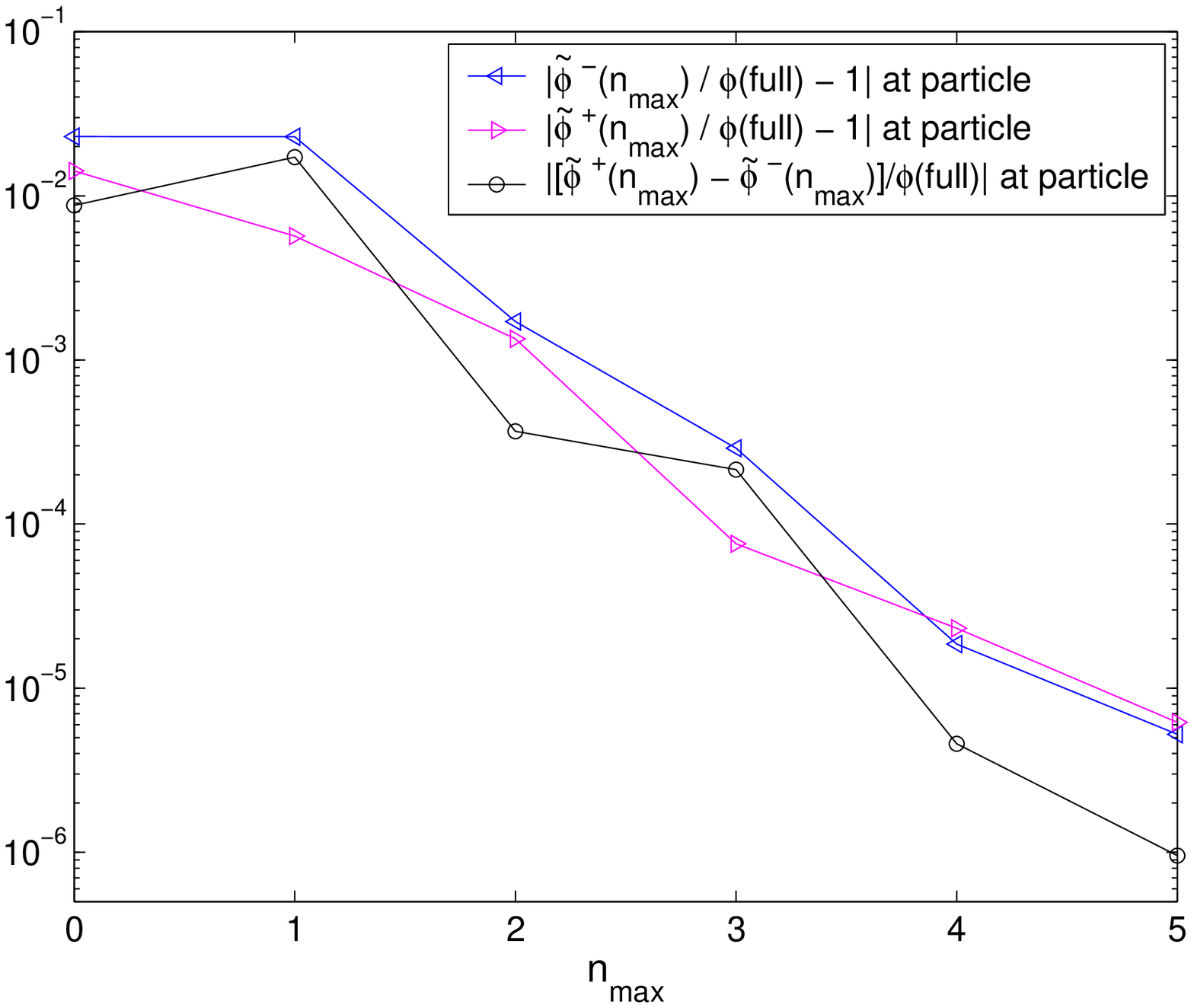}
            \epsfysize 7cm \epsfbox{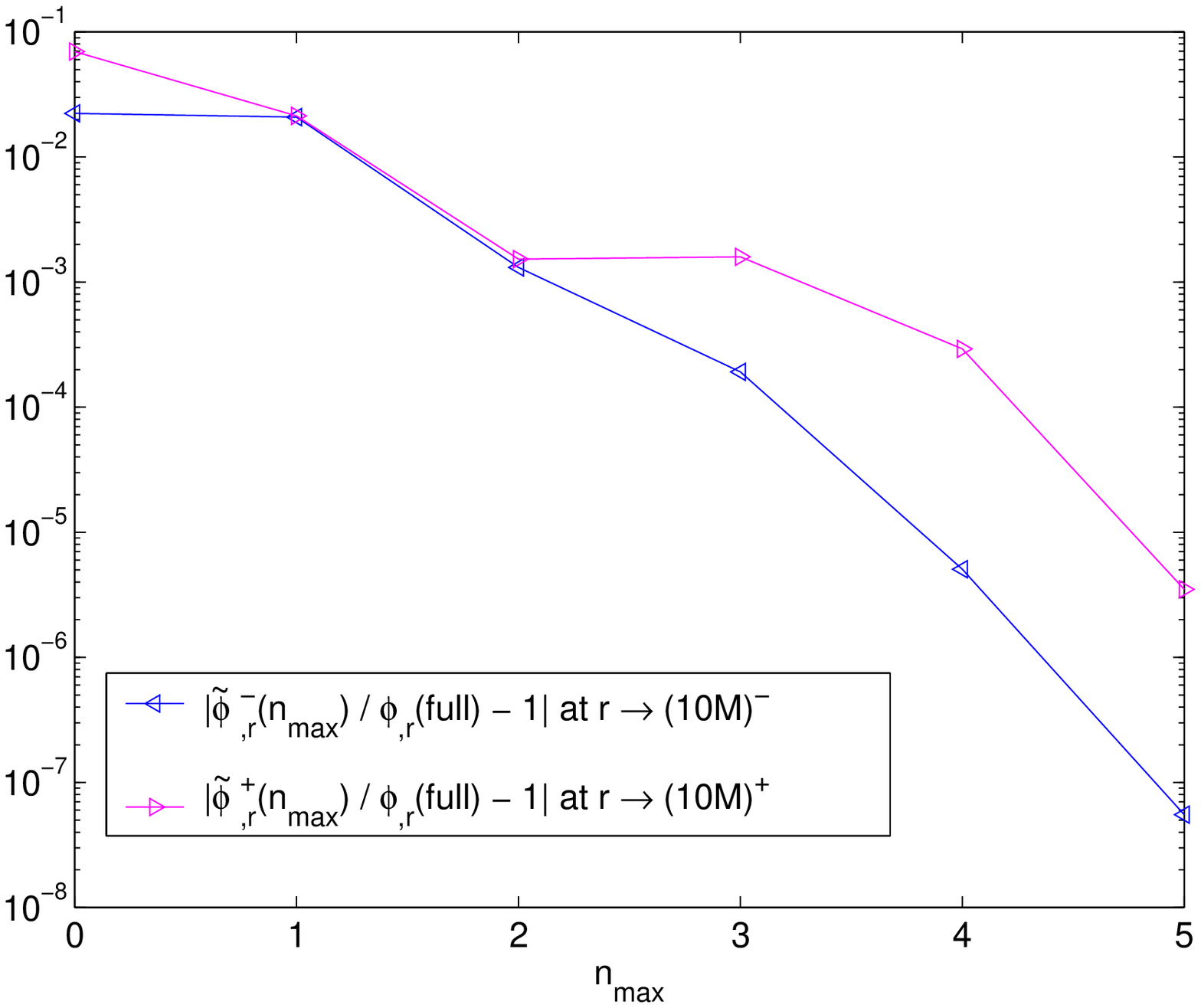}}
\caption{\protect\footnotesize
Convergence of the extended $n$-mode sum at the particle's location.
The left and right panels display the values {\em at the particle's location}
($r=10M$) of the fractional differences shown in the left and right panels
of Fig.\ {\ref{Fig:5}}, correspondingly.
The {\it left panel} demonstrates the exponential convergence of the
extended $n$-mode sum to the correct value of the field
at the particle. In particular, the mismatch between the partial sums for
$\tilde\phi^{+}$ and $\tilde\phi^{-}$ at the particle appears to vanish
exponentially with increasing $n_{\rm max}$.
As we suggest in Sec.\ \ref{Sec:conclusions}, one can in fact make a good
use of this mismatch in numerical calculations,
by recording its residual value and using it to assess the truncation error
of the partial $n$-mode sum.
Comparing with Figs.\ \ref{Fig:1} and \ref{Fig:3} (left panels), it is
striking that, in the example considered here, summing up to only $n=2$ with
the new method achieves a better accuracy in the local field than summing
over as many as $16$ modes using the standard approach.
The {\it right panel} demonstrates the fast convergence of the
derivatives $\tilde\phi_{,r}^{+}$ and $\tilde\phi_{,r}^{-}$ in the new
approach. An exponential convergence, expected form theory, is loosely
suggested from the data shown.
}
\label{Fig:6}
\end{figure}

\section{Discussion} \label{Sec:conclusions}

The basic relation underpinning our new method is expressed in Eq.\
(\ref{Eq:a117}), which describes the construction of the $lm$-multipole
of the scalar field from extended homogeneous solutions.
Eqsuation (\ref{Eq:Rtilde}) and (\ref{Eq:a125})--(\ref{Eq:a135}), in
conjunction with Eqs.\ (\ref{Eq:a105}) and (\ref{Eq:a45}), constitute
our new prescription for constructing the self force in the frequency domain.
The following list highlights the advantages of this new formulation.

\begin{itemize}
\item In the standard frequency-domain scheme, the Fourier sum over $\omega$-modes
suffers from Gibbs-type irregularities near the particle's location. In particular,
the Fourier sum fails to correctly recover the derivatives of the field's multipoles
at the particle (which are essential input in self-force calculations).
In the new method one constructs the local field's multipoles as Fourier sums of
globally homogeneous $\omega$-modes. These sums converge uniformly, circumventing
the above complication.
This is the essential and most crucial merit of the new method.
[Note also that the uniform convergence (which also applies to the {\em
derivatives} of the extended homogeneous mode functions) allows one to obtain the
derivatives of the field's multipoles using a term-by-term differentiation of the
individual Fourier components. This again leads to Eqs. (\ref{Eq:a130}) and
(\ref{Eq:a135}) above.]

\item
The sum over the homogeneous $\omega$ modes converges exponentially
everywhere, and even at the very location of the particle. This is extremely
convenient from a practical point of view. It should be noted that, within our
new scheme, it becomes ``as easy'' to obtain the local $l$-mode field near the
particle as it is to obtain the same field in the far-zone---in sharp contrast with the
situation in both the standard frequency-domain method and the time-domain method.

\item
In the standard scheme, each of the inhomogeneous $\omega$-modes
is computed via Eq.\ (\ref{Eq:a70}).
In practice, this involves the (numerical) evaluation of two integrals for each
value of $r$ between $r_{\rm min}$ and $r_{\rm max}$.
In the new scheme, the extended homogeneous
$\omega$-modes are obtained via Eq.\ (\ref{Eq:Rtilde}), which requires the same
two integrals (\ref{Eq:a80}) for {\em all} values of $r$. Thus, remarkably,
the new scheme does not only perform better mathematically---it is also much
simpler to implement.

\item
Finally, within the new scheme one is offered a convenient handle with which to monitor
and control the large-$\omega$ truncation error (i.e., the error caused by omission of the
terms $|n|>n_{\rm max})$: The residual amount by which the partial sum
$\sum_{n=-n_{\rm max}}^{n_{\rm max}}$ (for any of the quantities $\phi_{lm}^{i}$)
fails to be consistent with the appropriate jump condition at $r=r_{\rm p}(t)$
is a faithful measure of the truncation error. One can therefore conveniently keep the
latter below a set level by setting a threshold on the amount of residual inconsistency.

\end{itemize}

What is the application scope of the new method? In this paper we introduced the method
as applied specifically for scalar-field perturbations on the Schwarzschild background.
However, the basic idea is equally applicable to a wide range of other problems. It is
clear from our discussion that the problematics of reconstructing the local multipole field
as a sum over frequency modes has little bearing on the precise form of the
underlying field equations. Rather, it is a feature of the spherical-harmonic
decomposition: When we consider an individual $lm$ mode, we effectively convert the
physical problem of a point particle moving in an eccentric orbit into a problem of a
source ``shell'' (which expands and contracts over time); the perturbation field is
not smooth across the shell, which gives rise to Gibbs-type complications when we attempt
to express it as a sum over frequency modes. The same problem would occur in essentially
any perturbation treatment in Schwarzschild which incorporates a spherical harmonic
decomposition, such as the Teukolsky formalism (for EM and gravitational
perturbations), the standard Regge--Wheeler/Zerilli/Moncrief formalism of gravitational
perturbations, or the more direct Lorenz-gauge formulation
\cite{Barack:2005nr}. The method we propose here as a cure for the problem is directly
applicable for any of these treatments. A forthcoming paper \cite{BGSprep} will report on the
calculation of the monopole and dipole modes of the Lorenz-gauge metric perturbation
(for eccentric orbits in Schwarzschild), facilitated by the method of extended homogeneous solutions.
This calculation is now being incorporated in a code which computes the total
gravitational self-force for generic orbits in Schwarzschild \cite{SBprep}.

To what extent is our new method relevant for Kerr perturbations? Here the
situation is more subtle. The original mode-sum scheme for the self force \cite{Barack:2002mh}
(which sets the main context for the current work) incorporates a decomposition in
spherical harmonics even in the Kerr spacetime. This is, of course, technically possible
(see the footnoted remark at the Introduction),
although the resulting field equations then couple between different $l$-modes.
Regardless of the latter fact, each of the $lm$-modes in this decomposition would again
be sourced by an expanding/contracting thin shell, the non-smoothness of the perturbation across
this shell would give rise to the Gibbs phenomenon, and our method would provide an efficient
cure.

The situation is different if one tackles the Kerr problem by means of the
more natural decomposition in {\it spheroidal} harmonics, which decouples
the field equation in the frequency domain. Since the spheroidal-harmonic
functions depend on the frequency, one no longer has a strict notion of
a time-domain `$lm$ mode' in this case. One may (somewhat artificially) define
the ``spheroidal-harmonic $l'm$-mode'' $\Phi_{l'm}(t,r,\theta,\varphi)$
by summing over all $\omega$ for given spheroidal-harmonic numbers $l',m$.
However, in this case the effective geometric picture of a thin source
shell would no longer apply: In the procedure of Fourier decomposition of
the original point source (to obtain the source's $l' m \omega$ modes)
followed by a Fourier summation over $\omega$, the extra dependence of the
spheroidal harmonics on $\omega$ will cause the reconstructed source's $l' m$
mode function to deviate from a $\delta$-function in $t$ (for given
$r,\theta,\varphi$). Correspondingly, at a given $t$ the source's $l' m$
mode function will most likely represent a ``smeared'' shell.

Based on the above discussion one might conclude that the high-frequency
problem would not occur in the first place if spheroidal harmonics were
used (as in this case there would be no $\delta$-type shell).
We believe, however, that this may represent a false logic.
The viability of the mode-sum approach relies crucially on the fact that
the individual mode contributions $F_{l \mu}^{ \pm }$ in Eq.\ (\ref{Eq:a45})
are well-defined quantities. This fact is a direct consequence of the
perfect one-sided smoothness of the mode functions $\phi_{lm}(t,r)$ even
at the limit $r\to r_{\rm p}(t)$. This smoothness, in turn, stems from
the fact that for each spherical-harmonic $l,m$ the source term is
confined to a $\delta$-function over a shell---and this $\delta$-function
is distributed over the shell in a perfectly smooth manner.
Note also that the functions $\phi_{lm}(t,r)$ are {\it homogeneous}
time-domain solutions on both sides of this shell---even at the immediate
particle's neighborhood. In spheroidal-harmonic decomposition this
situation is changed, and the spheroidal-harmonic mode function
$\Phi_{l'm}(t,r,\theta,\varphi)$ defined above will no longer be a
homogeneous solution in the very neighborhood of the particle. It is
conceivable that in the immediate particle's neighborhood the smeared
source shell will be dominated by large-$\omega$ modes.
[The spherical-harmonics decomposition is protected against this
potential problem thanks to the combination of (i) the perfect
off-shell homogeneity, and (ii) the independence of the $l,m$
harmonic---and hence of the $\delta$-function distribution over the
spherical shell---on $\omega$.] Thus, in a spheroidal-harmonic
decomposition, the high-frequency problem may take a much more severe form:
It may endanger the very existence of $F_{l' \mu}^{ \pm }$ (namely the
spheroidal-harmonics analogs of $F_{l \mu}^{ \pm }$) as regular quantities,
which would in turn render the (spheroidal-harmonics analog of the)
mode-sum formula (\ref{Eq:a45}) meaningless.

The morphology of the spheroidal-harmonics smeared source shell still needs be
investigated, and especially its structure near the point charge.
The mode contributions $F_{l' \mu}^{ \pm }$
may turn out to be well-defined after all, but this is far from obvious.
In any case, a spheroidal-harmonics-based variant of the mode-sum method for Kerr
has not been developed yet to the best of our knowledge.
The existing Kerr mode-sum variant \cite{Barack:2002mh} incorporates the
spherical-harmonics decomposition, and as such it exhibits the same
high-frequency problem as in the Schwarzschild case. The method of extended
homogeneous solutions then elegantly resolves this problem in the Kerr case as well.

\section*{ACKNOWLEDGEMENTS}

LB and NS acknowledge support from PPARC/STFC through grant number PP/D001110/1.

\appendix

\section{Eccentric geodesics in Schwarzschild}
\label{AppA}

This appendix reviews the standard description of eccentric geodesics
in Schwarzschild spacetime, and provides the necessary formulas for all
orbital quantities (frequency, four-velocity, etc.) needed for the numerical
computations in Secs.\ \ref{Sec:Problem} and \ref{Sec:Solution}.

As in the main text, we consider a pointlike test particle in a bound equatorial
geodesic orbit around a Schwarzschild black hole with mass parameter $M$. The radial
location of the particle is bounded in the range $r_{\rm min}\leq r
\leq r_{\rm max}$, for some $r_{\rm max}>r_{\rm min}>4M$.
Such geodesics constitute a two-parameter family. Each geodesic
is uniquely characterized, for example, by the two ``turning point'' values
$r_{\rm min}$ and $r_{\rm max}$.
An alternative parameterization (originally due to Darwin \cite{Darwin})
employs the ``semi-latus rectum'', $p$, and ``eccentricity'', $e$, both
analogous to their counterparts from Keplerian celestial mechanics.
The parameter pairs $(p,e)$ and $(r_{\rm min},r_{\rm max})$ are related through
\begin{equation} \label {Eq:52}
p=\frac{2r_{\rm max}r_{\rm min}}{r_{\rm max}+r_{\rm min}}, \quad\quad
e=\frac{r_{\rm max}-r_{\rm min}}{r_{\rm max}+r_{\rm min}},
\end{equation}
or, inverting,
\begin{equation} \label {Eq:54}
r_{\rm max}=\frac{p}{1-e}, \quad\quad
r_{\rm min}=\frac{p}{1+e}.
\end{equation}

With the parameterization $(p,e)$, the orbital radius is given by
\begin{equation} \label {Eq:60}
r_{\rm p}(\chi)=\frac{p}{1+e\cos\chi},
\end{equation}
where $\chi$ is a monotonically-increasing parameter (``radial phase'')
along the worldline.
This parameter is related to the Schwarzschild time $t_{\rm p}$ along
the worldline through
\begin{equation} \label {Eq:55}
\frac{d\chi}{dt_{\rm p}}=\frac{(p-2M-2Me\cos\chi)(1+e\cos\chi)^2}{p^2}
\left(\frac{p/M-6-2e\cos\chi}{(p/M-2)^2-4e^2}\right)^{1/2},
\end{equation}
with the constant of integration fixed such that $\chi=0$ at some
``periastron'' passage ($r_{\rm p}=r_{\rm min}$). $\chi$ is related to the proper
time along the eccentric geodesic through
\begin{equation} \label {Eq:57}
\frac{d\chi}{d\tau}=\frac{(1+e\cos\chi)^2}{M(p/M)^{3/2}}
\left(\frac{p/M-6-2e\cos\chi}{p/M-3-e^2}\right)^{1/2}.
\end{equation}
The radius $r_{\rm p}$ is manifestly periodic in $\chi$, with $t$-period
\begin{equation} \label {Eq:65}
T_r\equiv \int_0^{2\pi}(d\chi/dt_{\rm p})^{-1}d\chi
\end{equation}
and radial frequency $\Omega_{r}\equiv 2\pi/T_{r}$.

For our numerical implementation of Eqs.\ (\ref{Eq:90}) and (\ref{Eq:C20})
(in App.\ \ref{AppD}), we start by choosing $r_{\rm min}$ and
$r_{\rm max}$ and
then use Eq.\ (\ref{Eq:52}) to determine $p$ and $e$ [or, alternatively, we
pick $p$ and $e$, then use Eq.\ (\ref{Eq:54}) to determine $r_{\rm min}$
and $r_{\rm max}$]. We then solve for $t_{\rm p}(\chi)$ for $0\leq\chi\leq\pi$
by integrating the inverse of Eq.\ (\ref{Eq:55}) numerically [taking $t_{\rm p}(0)=0$].
We hence obtain the radial period $T_r=2t_{\rm p}(\chi=\pi)$ and the radial
frequency $\Omega_{r}$. Using Eq.\ (\ref{Eq:60}) to express $\chi$ in terms
of $r$ along the orbit (again for $0\leq\chi\leq\pi$), we then obtain
$t_{\rm p}(\chi(r))\equiv t_{\rm p}(r)$, and, inverting in the range
$0\leq t_{\rm p}\leq T_r/2$, also $r_{\rm p}(t)$. Finally, $u^t$ is obtained as
a function of $t$ by writing $u^t=(dt_{\rm p}/d\chi)(d\chi/d\tau)$,
substituting from Eqs.\ (\ref{Eq:55}) and (\ref{Eq:57}), and then using
Eq.\ (\ref{Eq:60}) to express $\chi$ in terms of $r_{\rm p}(t)$. This procedure
yields all necessary orbital parameters and functions for our numerical
examples.

Since the numerical illustrations of this work focus on the monopole
mode, which is axially-symmetric, they do not require an explicit computation
of the azimuthal frequency $\Omega_{\varphi}$. For completeness, though, we
mention that this frequency is defined, in an orbit-average manner, by
\begin{equation} \label {Eq:17}
\Omega_{\varphi}\equiv \frac{1}{T_r}\int_0^{T_r}(d\varphi_{\rm p}/dt)dt.
\end{equation}
The ``local'' frequency, which depends on $r_{\rm p}$, is given in terms of
the $r$-phase $\chi$ as
\begin{equation} \label {Eq:18}
\frac{d\varphi_{\rm p}}{dt}=
\frac{(p/M-2-2e\cos\chi)(1+e\cos\chi)^2}{M(p/M)^{3/2}
[(p/M-2)^2-4e^2]^{1/2}}.
\end{equation}
The frequency $\Omega_{\varphi}$ can be computed by changing the integration
variable in Eq.\ (\ref{Eq:17}) to $\chi$, using Eq.\ (\ref{Eq:55}).

\section{Numerical construction of the homogeneous solutions $R_n^{\pm}(r)$}
\label{AppB}

We describe here the numerical construction of the frequency-domain homogeneous
basis ($R_n^{+}$, $R_n^{-}$) in the example of the scalar-field monopole.
For $l=0$, the homogeneous part of the $n$-mode radial ODE (\ref{Eq:a60})
takes the form
\begin{equation} \label {Eq:76}
\frac{d^2R_n^{\pm}}{dr_*^2}
+\left[n^2\Omega_r^2-2Mf(r)r^{-3}\right]R_n^{\pm}=0.
\end{equation}
For each $n$, $R_n^{+}$ and $R_n^{-}$ are independent solutions of this equation,
satisfying physical boundary conditions at $r\to\infty$ and $r\to 2M$, respectively.
These homogeneous solutions are needed both in constructing the `true' inhomogeneous
$n$-mode solution $R_n^{\rm inh}$ [via Eq.\ (\ref{Eq:90}) in App.\ \ref{AppD}]
and in constructing the extended homogeneous solutions $\tilde R^{\pm}_n$ within our
new method [via Eqs.\ (\ref{Eq:Rtilde}) and (\ref{Eq:C20})].

Consider first the static mode, $n=0$, which can be solved for analytically:
We have, simply, $R_0^-=r$ and $R_0^+=r\ln f$, which constitutes a unique basis
(up to multiplicative constants) with the property that $R_0^-$ is regular at the
event horizon and $R_0^+$ is regular at $r\to\infty$. [To see this, recall
that the static mode of the actual, full scalar field is $\Phi\propto R^{\pm}_0/r$,
with corresponding ``internal'' solution $\Phi^-\propto{\rm const}(\ne 0)$
and ``external'' solution $\Phi^+\propto\ln f$. The former is regular
at the horizon but fails to vanish at infinity, while the latter falls off
as $1/r$ at infinity but diverges at the horizon.]

For each mode $n\ne 0$ we solve Eq.\ (\ref{Eq:76}) numerically with suitable
boundary conditions, as we now describe. Our numerical domain is a one-dimensional
array representing physical radii in the range $r_{\rm in}\leq r\leq r_{\rm out}$.
The boundaries are taken to lie in the asymptotic vacuum domains:  $r_{\rm in}/(2M)-1\ll 1$
(the event horizon) and $r_{\rm out}\gg M$ (spatial infinity). In practice,
it proved sufficient for our purpose to set $r_{\rm in}=2.001M$ and
$r_{\rm out}=1000M$.

Consider first $R_n^-$. As an inner boundary condition for this function
we use the ansatz
\begin{equation}\label{Eq:A10}
R_n^-(r)=
e^{-in\Omega_r r_*}\sum_{k=0}^{k^-_{\rm max}}a_{nk}^-(r-2M)^{k}
\quad \text{(at $r=r_{\rm in}$)},
\end{equation}
where $a_k^-$ are coefficients to be determined below, and $k^-_{\rm max}$ is
taken large enough to guarantee that truncation error is kept below a prescribed
threshold (in practice, $k^-_{\rm max}=10$ proved sufficient for the values of $e,p,n$
considered in this work). The oscillatory factor in Eq.\ (\ref{Eq:A10}) is chosen such
that the contribution from each $n$-mode to the full monopole field attains the asymptotic
form $\propto\exp[-in\Omega_r(t+r_*)]$ as $r\to 2M$ [recall Eq.\ (\ref{Eq:72})].
This represents purely {\em ingoing} radiation, which is the correct physical condition at
the event horizon. To determine the coefficients $a^-_{nk}$ we substitute
Eq.\ (\ref{Eq:A10}) in the field equation (\ref{Eq:76}) and solve the resulting
hierarchy of algebraic equations at each order in $r-2M$.
This yields the following recursion formula for the coefficients $a^-_{nk}$
(with given $n$; we omit here the index $n$ for brevity):
\begin{eqnarray}\label{Eq:A20}
a_{k>0}^-&=&\frac{-1}{2M^2k( k -4i M n \Omega_r)}\times
\left\{
\left[M (2 k - 1) (k - 2) -12 i M^2 n \Omega_r (k - 1)\right]a_{k - 1}^-
\right.
\nonumber\\
&& \left.
+\left[(k - 3)/2 - 6 i M n \Omega_r\right] (k - 2)a_{k - 2}^-
- i n \Omega_r (k - 3) a_{k - 3}^-\right\},
\end{eqnarray}
with $a_{k<0}^-=0$. All coefficients $a^-_{k>0}$ are constructed recursively
given $a_0^-$, and are all proportional to $a_0^-$.

To solve Eq.\ (\ref{Eq:76}) for $R_n^-(r)$, we simply set $a_0^-=1$, impose the values
of $R_n^-$ and $dR_n^-/dr$ at $r_{\rm in}$ using Eq.\ (\ref{Eq:A10}), and integrate
numerically forward from $r=r_{\rm in}$ to $r=r_{\rm max}$ (the value of $R_n^-$
at $r>r_{\rm max}$ is not needed in our analysis).

Now consider $R_n^+$. As an outer boundary condition for this function we take
\begin{equation}\label{Eq:A30}
R_n^+(r)=e^{in\Omega_r r_*}\sum_{k=0}^{k^+_{\rm max}}a_{nk}^+\,r^{-k}
\quad \text{(at $r=r_{\rm out}$)},
\end{equation}
where $a_{nk}^+$ are determined below, and $k^+_{\rm max}$ is chosen, once again,
such that truncation error is kept below a set threshold (here, too, $k^+_{\rm max}=10$
was sufficient in our analysis). With this condition, the contribution from each
$n$-mode to the full monopole field has the asymptotic form $\propto\exp[-in\Omega_r(t-r_*)]$
as $r\to\infty$, representing purely {\em outgoing} radiation---the correct physical
condition at infinity. The expansion coefficients in Eq.\ (\ref{Eq:A30}) are obtained
recursively using
\begin{equation}\label{Eq:A40}
a_{k>0}^+ = \frac{i}{2n \Omega_r k}
\left[-k(k - 1)a_{k - 1}^+ + 2M (k - 1)^2 a_{k - 2}^+ \right],
\end{equation}
with $a_{k<0}^+=0$. (Recall $n=0$ is dealt with analytically, so one need not
worry about the ill-definiteness of the recursion relation in this case.)
All coefficients $a^+_{k>0}$ are constructed recursively given $a_0^+$, and are all
proportional to $a_0^+$.

To solve Eq.\ (\ref{Eq:76}) for $R_n^+$, we set $a_0^+=1$,
impose the values of $R_n^+$ and $dR_n^+/dr$ at $r_{\rm out}$ using Eq.\ (\ref{Eq:A30}),
and integrate numerically backward from $r=r_{\rm out}$ to $r=r_{\rm min}$ (the value of
$R_n^+$ at $r<r_{\rm min}$ is not needed in our analysis).

\section{Construction of the inhomogeneous solutions $R_n^{\rm inh}$}
\label{AppD}

This Appendix details the computation of the inhomogeneous radial functions
$R_n^{\rm inh}(r)$ for our numerical illustration in Sec.\ \ref{Subsec:old}.

For $l=0$, the functions $R^{\rm inh}_n(r)$ satisfy the ODE
\begin{equation} \label {Eq:74}
\frac{d^2R^{\rm inh}_n}{dr_*^2}
+\left[n^2\Omega_r^2-2Mf(r)r^{-3}\right]R^{\rm inh}_n=Z_n,
\end{equation}
where the source term is obtained by taking the inverse-Fourier
transform in (the discrete, $l=0$ version of) Eq.\ (\ref{Eq:a55}):
\begin{eqnarray} \label {Eq:75}
Z_n(r)&=&T_r^{-1}\int_0^{T_r}\left[-4\pi r f(r)\hat\rho_{l=m=0}(t,r)\right]
e^{in\Omega_r t}dt
\nonumber\\
&=& -q(4\pi)^{1/2} T_r^{-1} \int_0^{T_r} f(r)(r u^t)^{-1}
\delta[r-r_{\rm p}(t)] e^{in\Omega_r t}dt.
\end{eqnarray}
Here $T_r$ is the radial period [see Eq.\ (\ref{Eq:65})],
and in the second equality we have substituted for $\hat\rho$ from Eq.\
(\ref{Eq:a22}), setting $m=0$ and $\hat c_{l=m=0}=(4\pi)^{-1/2}$.
Note in Eq.\ (\ref{Eq:75}) that $Z_n=0$ for $r<r_{\rm min}$ or
$r> r_{\rm max}$, and that for any $r_{\rm min}< r< r_{\rm max}$ the
integrand is supported only at two points within the integration domain---the two
times $t$ for which $r_{\rm p}(t)=r$.
Changing the integration variable from $t$ to $r_{\rm p}$ one thus obtains
\begin{equation} \label {Eq:82}
Z_n(r)=
-\frac{2q(4\pi)^{1/2} f(r)}{T_r r|u^r(r)|}\cos[n\Omega_n t_{\rm p}(r)]
\times\Theta(r-r_{\rm min})\times\Theta(r_{\rm max}-r),
\end{equation}
where $\Theta$ is the standard unit step function, $u^r=dr_{\rm p}/d\tau$
is the $r$ component of the particle's four-velocity, and $t_{\rm p}(r)$
is the result of inverting $r=r_{\rm p}(t)$ in the domain
$0\leq t\leq T_r/2$, assuming $r_{\rm p}(0)=r_{\rm min}$
[note $r_{\rm p}(t)$ is single-valued in this restricted domain].

In the standard approach, the physical solution to Eq.\ (\ref{Eq:74}) is
obtained through the formula (\ref{Eq:a70}), replacing
$R_{lm\omega}\to R^{\rm inh}_{n}$, $R_{lm\omega}^{\pm}\to R_{n}^{\pm}$,
and $Z_{lm\omega}\to Z_n$.
Here $R_n^{+}$ and $R_n^{-}$
are two independent solutions to the homogeneous part of Eq.\ (\ref{Eq:74}),
satisfying physical boundary conditions at $r\to\infty$ and $r=2M$,
respectively. It is convenient to change the integration variable in
Eq.\ (\ref{Eq:a70}) from $r$ to $t_{\rm p}(r)$ [taking $t_{\rm p}(r_{\rm min})=0$],
which avoids the singularity in $Z_n(r)$ at $r=r_{\rm max},r_{\rm min}$.
Substituting for $Z_n$ from Eq.\ (\ref{Eq:82}), and using
$|u^r|=u^r=u^t(dr_{\rm p}/dt)$ for $0\leq t\leq T_r/2$,
Eq.\ (\ref{Eq:a70}) becomes
\begin{eqnarray} \label {Eq:90}
R^{\rm inh}_n(r)=-2q(4\pi)^{1/2}T_r^{-1}W^{-1}&\times&\left[
R_n^+(r)\int_{0}^{{\hat t}_{\rm p}(r)}\frac{R_n^-(r_{\rm p}(t))}
{r_{\rm p}(t) u^t(r_{\rm p}(t))}\cos(n\Omega_r t)dt   \right.
\nonumber\\
&&+
\left.
R_n^-(r)\int_{{\hat t}_{\rm p}(r)}^{T_r/2}\frac{R_n^+(r_{\rm p}(t))}
{r_{\rm p}(t) u^t(r_{\rm p}(t))}\cos(n\Omega_r t)dt
\right],
\end{eqnarray}
where we have introduced
\begin{equation} \label {Eq:100}
{\hat t}_{\rm p}(r)=\left\{
\begin{array}{ll}
0,               & r\leq r_{\rm min}, \\
t_{\rm p}(r),    & r_{\rm min}\leq r\leq r_{\rm max}, \\
T_r/2,           & r\geq r_{\rm max}.
\end{array}
\right.
\end{equation}
The solution $R^{\rm inh}_n$ in Eq.\ (\ref{Eq:90}) is manifestly an
analytic function of $r$ everywhere, except at $r=r_{\rm min},r_{\rm max}$.
Note, recalling Eq.\ (\ref{Eq:a75}), that {\em outside} the domain
$r_{\rm min}<r<r_{\rm max}$ Eq.\ (\ref{Eq:90}) reduces to the homogeneous
solutions
\begin{equation} \label {Eq:C10}
R_{n}^{\rm inh}(r)=\left\{
\begin{array}{ll}
C_{n}^- R_{lm\omega}^-(r) ,
& r\leq r_{\rm min}, \\
\\
C_{n}^+ R_{lm\omega}^+(r) ,
& r\geq r_{\rm max},
\end{array}
\right.
\end{equation}
where the coefficients $C_{n}^\pm$ are given by
\begin{eqnarray} \label {Eq:C20}
C^{\pm}_n=-2q(4\pi)^{1/2}T_r^{-1}W^{-1} \,
\int_{0}^{T_r/2}\frac{R_n^{\mp}(r_{\rm p}(t))}
{r_{\rm p}(t) u^t(r_{\rm p}(t))}\cos(n\Omega_r t)dt.
\end{eqnarray}

Equation (\ref{Eq:90}) can be implemented numerically to obtain
$R^{\rm inh}_n(r)$, in the following manner:
We specify the physical orbit by picking the values of $r_{\rm min}$ and
$r_{\rm max}$, and then use the relations given in App.\ \ref{AppA} to obtain
(numerically) the values of $\Omega_r$ and $T_r$ and the functions $r_{\rm p}(t)$,
$t_{\rm p}(r)$ and $u^t(r_{\rm p}(t))$ for the specified orbit.
We next construct the homogeneous basis $R_n^{\pm}(r)$ by numerically integrating
the homogeneous part of Eq.\ (\ref{Eq:74}) with suitable boundary conditions.
This procedure is described in
App.\ \ref{AppB}. Once the solutions $R_n^{\pm}(r)$ are at hand, the (constant)
value of the Wronskian is obtained using Eq.\ (\ref{Eq:aw}). Finally, for each given $n$,
we calculate the integrals in Eq.\ (\ref{Eq:90}) numerically, and construct the
solution $R^{\rm inh}_n(r)$.

\section{High-frequency analysis}
\label{AppC}

In this appendix we analyze the convergence and analyticity of the Fourier sum
involved in the construction of the key functions $\phi_{lm}^{(i)}$ in the
``extended homogeneous solutions'' approach---i.e., the sum over $n$ on
the right-hand side of Eq.\ (\ref{Eq:a116}). Both convergence and analyticity
are crucial for the definiteness and validity
of our proposed approach. By analyzing the behavior of the mode sum in the
high-frequency limit, we provide here a strong indication this sum
converges (at least) exponentially in the entire domain $r>2M$, and is
therefore also analytic in this entire domain.

\subsection{WKB approximation for the large-$\omega $ homogeneous radial
functions}

The extended frequency-domain radial functions are give by
\begin{equation}
{\tilde{R}}_{lm\omega }^{\pm }(r)=C_{lm\omega }^{\pm }R_{lm\omega }^{\pm
}(r) \quad (r>2M),
\label{ap10}
\end{equation}
where the ($r$-independent) coefficients $C_{lm\omega }^{\pm }$ are given
in Eq.\ (\ref{Eq:a80}),
and $\{R_{lm\omega }^{+}(r),R_{lm\omega }^{-}(r)\}$ is a pair of independent
solutions to the homogeneous equation
\begin{equation} \label {aa5}
\frac{d^2R_{lm\omega}}{dr_*^2}
+\left[\omega^2-V_l(r)\right]R_{lm\omega}=0,
\end{equation}
satisfying suitable boundary conditions at $r\to\infty$ and $r\to 2M$, respectively.
In what follows we shall explore the asymptotic behavior of the quantities
${\tilde{R}}_{lm\omega }^{\pm }$ at large $\omega$, and consequently evaluate the
large-$\omega$ contribution to the extended time-domain functions
$\tilde\phi_{lm}^{\pm}(t,r)$ in Eq.\ (\ref{Eq:a125}).

Since the potential $V_{l}(r)$ is bounded (for a given $l$), in the large-$
\omega $ limit the term in square brackets in Eq.\ (\ref{aa5}) is dominated
by $\omega ^{2}$. Using the WKB approximation, we can then write
\begin{equation}
R_{lm\omega }^{\pm }\cong \left[ 1-V_{l}(r)/\omega ^{2}\right]
^{-1/4}\exp\left( \pm i\int \sqrt{\omega ^{2}-V_{l}(r)}dr_{*}\right).  \label{aa10}
\end{equation}
The square-root term may be expanded for large $\omega $ as
\begin{equation}
\sqrt{\omega ^{2}-V_{l}(r)}\cong \omega -V_{l}(r)/(2\omega).
\end{equation}
We shall only consider here the leading-order solution at large $\omega $,
so we ignore the term $V_{l}(r)/(2\omega) $ in this expression, as well as the
$\propto \omega ^{-2}$ term in the square brackets in Eq.\ (\ref{aa10}). We obtain
\begin{equation}
R_{lm\omega }^{\pm }\cong e^{\pm i\omega r_{*}}.  \label{aa20}
\end{equation}
From the form of the ODE (\ref{aa5}) it is clear that the Wronskian $W$ is constant,
and for the specific pair $R_{lm\omega }^{\pm }$ in Eq.\ (\ref{aa20}) it takes the value
\begin{equation}
W=2i\omega . \label{aaW}
\end{equation}

\subsection{The source term $Z_{lm\omega}$ and its spectrum}

To calculate the coefficients $C_{lm\omega }^{\pm }$ we first need to
analyze the source term $Z_{lm\omega }(r)$ and determine its discrete
spectrum. Inverting the Fourier transform in Eq.\ (\ref{Eq:a55}), we have
\begin{equation}
Z_{lm\omega }(r)=f^{[1]}(r)\int _0^{T_r} \hat{\rho}_{lm}(t,r)e^{i\omega t}dt,\quad
\label{aa15}
\end{equation}
where $T_r$ denotes the $t$-period of the radial motion (see App.\ \ref{AppA}).
Throughout this appendix, $f^{[k]}$ ($k=1,2,3,...$) denote certain functions of $r$,
independent of $\omega $ (or $n$), which are {\em analytic} throughout $r>2M$
(but whose precise form would not interest us otherwise).

Our first goal is to obtain the $\omega $ spectrum of $\hat{\rho}_{lm}$.
For an eccentric geodesic the azimuthal motion may be expressed as
\begin{equation}
\varphi_{\rm p}(t)=\Omega_{\varphi} t+\Delta \varphi(t),
\end{equation}
where $\Omega_{\varphi}$ is the
$t$-averaged angular frequency $d\varphi /dt$ (see App.\ \ref{AppA}), and
$\Delta \varphi(t)$ is a certain analytic function of $t$, which is periodic with periodicity $T_r$.
[Analyticity is directly inherited from that of $\varphi_{\rm p}(t)$;
The $T_r$-periodicity results from the fact that for a given orbit
$d \Delta \varphi/dt$ is a function of $r$ only.]
Substituting this in Eq.\ (\ref{Eq:a22}) we get
\begin{equation}
\hat{\rho}_{lm}(t,r)=\hat c_{lm}q(r^{2}u^{t})^{-1} \beta_{m}(t)
e^{-im\Omega_{\varphi} t}\delta [r-r_{\rm p}(t)],
\label{a22a}
\end{equation}
where $\beta_{m}(t)\equiv e^{-im\Delta \varphi(t)}$ is an analytic, $T_r$-periodic,
function of $t$.
We rewrite this as
\begin{eqnarray}
 \hat{\rho}_{lm}(t,r)&=& \left(  q f^{[2]}_{lm}(r) \beta_{m}(t) \delta[r-r_{\rm p}(t)] \right)
e^{-im\Omega_{\varphi} t}
\nonumber\\
&\equiv&
S_{lm}(t,r)e^{-im\Omega_{\varphi} t}.
\label{aa30}
\end{eqnarray}
The spectrum of $\hat{\rho}_{lm}$ is the same as that of $S_{lm}$, with all
frequencies simply shifted by $m\Omega_{\varphi}$.
Now, the dependence of $S_{lm}$ on $t$
is only through the $T_{r}$-periodic functions $r_{\rm p}(t)$ and $\beta_{m}(t)$.
Therefore we may write
\begin{equation}
S_{lm}(t,r)=\sum_{n}S_{lmn}(r)e^{-in\Omega _{r}t},
\end{equation}
and correspondingly
\begin{equation}
\hat{\rho}_{lm}(t,r)=\sum_{n}S_{lmn}(r)e^{-i(m\Omega_{\varphi} +n\Omega _{r})t},
\end{equation}
where $\Omega _{r}= 2 \pi/T_{r}$ is the fundamental radial frequency.
Thus, the spectrum of $\hat{\rho}_{lm}(t,r)$ is the discrete set of frequencies
\begin{equation}
\omega =m\Omega_{\varphi} +n\Omega _{r}\equiv \omega _{nm} .
\label{aa33}
\end{equation}

Finally, we calculate the coefficients $Z_{lmn}(r)\equiv Z_{lm\omega _{nm}}(r)$
using Eq.\ (\ref{aa15}). Pulling the factor $f^{[1]}(r)$ into the integral
and using Eqs.\ (\ref{aa30}) and (\ref{aa33}), we get
\begin{equation}
Z_{lmn}(r)=q\int _0^{T_r} f_{lm}^{[3]}(r) \beta_{m}(t) \delta [r-r_{\rm p}(t)]e^{in\Omega
_{r}t}dt.
\label{aa45}
\end{equation}

\subsection{Calculating the coefficients $C^{\pm}_{lm\omega}$ }

We turn now to calculate the coefficients
$C_{lmn}^{\pm }\equiv C_{lm\omega_{nm}}^{\pm }$. These are given by
\begin{equation}
C_{lmn}^{\pm }=W^{-1}\int_{r_{\mathrm{min}}}^{r_{\rm max}}\frac{R_{lmn}^{\mp
}(r)Z_{lmn}(r)}{f(r)}\,dr,
\end{equation}
where $R_{lmn}^{\mp }\equiv R_{lm\omega _{nm}}^{\mp }$.
Substituting from Eq.\ (\ref{aa45}) for $Z_{lmn}$ and absorbing the factor $1/f(r)$
in $f_{lm}^{[3]}$ (to form another analytic function, $f_{lm}^{[4]}$) we get
\begin{equation}
C_{lmn}^{\pm }=qW^{-1}\int_{r_{\mathrm{min}}}^{r_{\rm max}}\int_0^{T_r}
 f_{lm}^{[4]}(r) R_{lmn}^{\mp }(r) \beta_{m}(t)
 \delta [r-r_{\rm p}(t)] e^{in\Omega _{r}t}dtdr .
\end{equation}
The two integrals in the last equation are allowed to be interchanged (as
one may verify, for example, by explicitly carrying out first the $t$-integral
and only then the $r$-integral). One obtains
\begin{eqnarray}
C_{lmn}^{\pm } =
qW^{-1}\int _0^{T_r} f_{lm}^{[4]}[r_{\rm p}(t)] R_{lmn}^{\mp
}[r_{\rm p}(t)] \beta_{m}(t)\,e^{in\Omega _{r}t}dt.
\end{eqnarray}
Next we substitute for $W$ from Eq.\ (\ref{aaW}), and also for $R_{lmn}^{\mp }$ using
the large-$\omega $ asymptotic form given in Eq.\ (\ref{aa20}).
Denoting $r_{*}^{\rm p}(t)\equiv r_{*}[r=r_{\rm p}(t)]$, we thus obtain
\begin{eqnarray*}
C_{lmn}^{\pm } &\cong &(-i/2)q\omega _{nm}^{-1}\int _0^{T_r}
f_{lm}^{[4]}[r_{\rm p}(t)] \beta_{m}(t) e^{\mp i\omega _{nm}r_{*}^{\rm p}(t)} e^{in\Omega _{r}t}dt \\
&=&(-i/2)q\omega _{nm}^{-1}\int _0^{T_r} f_{lm}^{[4]}[r_{\rm p}(t)]
\beta_{m}(t) e^{\mp im\Omega_{\varphi}
r_{*}^{\rm p}(t)} e^{in\Omega _{r}[t\mp r_{*}^{\rm p}(t)]}dt .
\end{eqnarray*}
The analytic factor $e^{\mp im\Omega_{\varphi}r_{*}^{\rm p}(t)}$ (as well as $-i/2$) may
be absorbed in $f_{lm}^{[4]}$---which will in turn become a new analytic function, denoted
$f_{lm}^{[5]\mp}$.
Defining
\begin{equation}
t_{\pm }\equiv t\pm r_{*}
\end{equation}
(namely the two Eddington--Finkelstein null coordinates), the last equation becomes
\begin{equation}
C_{lmn}^{\mp }\cong q\omega _{nm}^{-1}\int _0^{T_r} f_{lm}^{[5]\pm }[r_{\rm p}(t)]
\beta_{m}(t) e^{in\Omega_{r}t_{\pm }^{\rm p}(t)}dt ,
\label{ap30}
\end{equation}
where $t_{\pm }^{\rm p}(t)$ is $t_{\pm }(r=r_{\rm p}(t))\equiv t\pm r_{*}^{\rm p}(t)$.

In the next step we wish to transform the integration variable from $t$ to
$t_{\pm }^{\rm p}(t)$. To this end we briefly discuss the properties of this
transformation (and its inverse), particularly in terms of analyticity and periodicity.
Obviously $t_{\pm }^{\rm p}(t)$ is analytic.
Also, since the orbit is timelike, $t_{\pm }^{\rm p}(t)$ is monotonically increasing,
and $dt_{\pm }^{\rm p}/dt$ nowhere vanishes.
This implies that the inverse function $t(t_{\pm }^{\rm p})$ is well-defined and analytic.
Note also that all functions of $r_{\rm p}$ are periodic in $t_{\pm }^{\rm p}$,
with the same period $T_{r}$.
The same applies to $\beta_{m}(t)$.
Therefore Eq.\ (\ref{ap30}) may be expressed as
{\footnote{The integration limits in this expression are shifted, both by
$r_{*}^{\rm p}(t=0)$, with respect to those in Eq.\ (\ref{ap30}), but this
shift does not make any difference because the integration is still over a full period.}}
\begin{equation}
C_{lmn}^{\mp }\cong q\omega _{nm}^{-1}\int_0^{T_r} f_{lm}^{[5]\pm }[r_{\rm p}(t_{\pm }^{\rm p})]
\beta_{m}[t(t_{\pm }^{\rm p})] \,
\frac{dt}{dt_{\pm }^{\rm p}}e^{in\Omega _{r}t_{\pm }^{\rm p}}dt_{\pm }^{\rm p} .
\end{equation}

The last expression is nothing but $(q T_{r} \omega _{nm}^{-1})$ times the $n$-th
Fourier coefficient of the periodic function
\begin{equation}
H_{lmn}^{\pm }(t_{\pm }^{\rm p}) \equiv
 f_{lm}^{[5]\pm }[r_{\rm p}(t_{\pm }^{\rm p})]\,\beta_{m}[t(t_{\pm }^{\rm p})] \,\frac{dt}{dt_{\pm }^{\rm p}}.
\end{equation}
Since all three factors on the right-hand side are analytic functions of $t_{\pm }^{\rm p}$,
so is $H_{lmn}^{\pm }(t_{\pm }^{\rm p})$.
Therefore its Fourier coefficients must decay (at least) exponentially with $|n|$, and the same applies to the
coefficients $C_{lmn}^{\mp }$.

\subsection{Reconstructing the time-domain extended homogeneous solutions}

We turn now to explore the high-$\omega $ contribution to the time-domain
extended homogeneous solutions
\begin{equation}
\tilde{\phi}_{lm}^{\pm }(t,r)= \sum_{n}\tilde{R}_{lmn}^{\pm }(r)e^{-i\omega
_{nm}t}=\sum_{n}C_{lmn}^{\pm }R_{lmn}^{\pm }(r)e^{-i\omega _{nm}t}.\quad
\end{equation}
Using the large-$\omega $ asymptotic expression (\ref{aa20}) one finds
for the large-$\omega$ contribution
\begin{eqnarray}
\tilde{\phi}_{lm}^{\pm }(t,r)&\cong& \sum_{n}C_{lmn}^{\pm }e^{i\omega
_{nm}(-t\pm r_{*}^{\rm p})}=\sum_{n}C_{lmn}^{\pm }e^{-i\omega _{nm}t_{\mp}}
\nonumber\\
&=& e^{-im\Omega_{\varphi} t_{\mp }}\sum_{n}C_{lmn}^{\pm }e^{-in\Omega
_{r}t_{\mp}}.
\quad\quad\quad\label{ap50}
\end{eqnarray}
The sum in the last expression has the form of a Fourier series in time $t_{\mp}$,
with coefficients $C_{lmn}^{\pm }$ which, as we already established, decay at least
exponentially. Therefore, this sum converges to an analytic function of $t_{\mp}$.
It then follows that the entire right-most expression in Eq.\ (\ref{ap50}) is analytic
in $t_{\mp}$.
Since $t_{\mp}$, in turn, is an analytic function of $r$ and $t$, then so must be this
expression. We therefore conclude that, at least within the leading-order approximation,
{\em the large-$\omega$ contribution to $\tilde{\phi}_{lm}^{\pm}(t,r)$ is
analytic in both $r$ and $t$ throughout $r>2M$}. Note also that the sum over $n$
in Eq.\ (\ref{ap50}) is guaranteed to converge {\it uniformly} for all $t$
and $r>2M$. The same applies to the corresponding Fourier sums for the $r$ and $t$
derivatives of $\tilde\phi_{lm}^{\pm}$.

In the above discussion we considered only the leading-order term in the
$1/\omega$ WKB expansion. One naturally expects that the contribution from
higher-order terms in this expansion will converge even faster, and hence will
not interfere with the analyticity of $\tilde{\phi}_{lm}^{\pm}(t,r)$.
Also, such higher-order contributions are not expected to affect the uniform
convergence of the sum over extended $n$ modes.

We regard the results of the above leading-order calculation as a strong
indication that the extended time-domain solutions, as they are defined in
Eq.\ (\ref{Eq:a116}), are indeed analytic everywhere outside the black hole.
This is further supported by the numerical results presented in
Sec.\ \ref{Subsec:new}.

\end{document}